\documentclass[11pt]{article}
\usepackage{cite}
\usepackage[numbers,sort&compress]{natbib}

\usepackage{amsmath}
\usepackage{graphicx}
\usepackage{amsfonts}
\usepackage{bm}
\usepackage{amssymb}
\usepackage{epsfig}
\usepackage{color}
\usepackage{psfrag}
\usepackage{mathrsfs}
\usepackage{esint}
\usepackage{pifont}
\usepackage{amsthm}
\usepackage{url}
\usepackage{wasysym} 
\usepackage{nicefrac} 
\numberwithin{equation}{section}
\usepackage{stackengine}
\usepackage{comment}
\usepackage{stmaryrd,scalerel} 
\usepackage{subcaption}

\usepackage{tikz}
\usetikzlibrary{calc}
\usepackage{relsize}
\usepackage{bm}
\usepackage{moresize}
\usepgflibrary{shapes.geometric}
\usetikzlibrary{chains, positioning, quotes}
\usetikzlibrary{shapes,snakes}
\usetikzlibrary{snakes}
\usetikzlibrary{automata}
\usetikzlibrary{positioning}
\usetikzlibrary{arrows,decorations.pathmorphing,patterns,backgrounds,positioning,fit,petri}
\usetikzlibrary{automata}
\usetikzlibrary{graphs}
\usetikzlibrary{shadows}
\definecolor{colordiagram1}{RGB}{155, 255, 217}
\definecolor{colordiagram2}{RGB}{255, 148, 37} 
\definecolor{myyellow}{RGB}{255, 207, 0}

\def\dashint{\,\ThisStyle{\ensurestackMath{\stackinset{c}{.2\LMpt}{c}{.5\LMpt}{\SavedStyle-}{\SavedStyle\phantom{\int}}}\setbox0=\hbox{$\SavedStyle\int\,$}\kern-\wd0}\int}

\begin{document} \setcounter{page}{0}
\topmargin 0pt
\oddsidemargin 5mm
\renewcommand{\thefootnote}{\arabic{footnote}}
\newpage
\setcounter{page}{1}
\topmargin 0pt
\oddsidemargin 5mm
\renewcommand{\thefootnote}{\arabic{footnote}}
\newpage

\begin{titlepage}
\begin{flushright}
\end{flushright}
\vspace{0.5cm}
\begin{center}
{\large {\bf Correlations and structure of interfaces in the Ising model.\\Theory and numerics}}\\
\vspace{1.8cm}
{\large Alessio Squarcini$^{1,2,\natural}$ and Antonio Tinti$^{3,\flat}$ }\\
\vspace{0.5cm}
{\em $^1$Max-Planck-Institut f\"ur Intelligente Systeme,\\
Heisenbergstr. 3, D-70569, Stuttgart, Germany}\\
{\em $^2$IV. Institut f\"ur Theoretische Physik, Universit\"at Stuttgart,\\
Pfaffenwaldring 57, D-70569 
Stuttgart, Germany}\\
{\em $^3$Dipartimento di Ingegneria Meccanica e Aerospaziale,\\
Sapienza Universit\`a di Roma, via Eudossiana 18, 00184 Rome, Italy }\\
\end{center}
\vspace{1.2cm}

\renewcommand{\thefootnote}{\arabic{footnote}}
\setcounter{footnote}{0}

\begin{center}
\today
\end{center}

\begin{abstract}
We consider phase separation on the strip for the two-dimensional Ising model in the near-critical region. Within the framework of field theory, we find exact analytic results for certain two- and three-point correlation functions of the order parameter field. The analytic results for order parameter correlations, energy density profile, subleading corrections and passage probability density of the interface are confirmed by accurate Monte Carlo simulations we performed.
\end{abstract}

\vfill
$^\natural$squarcio@is.mpg.de, $^\flat$antonio.tinti@uniroma1.it
\end{titlepage}

\newpage
\tableofcontents
\newpage

\section{Introduction}
\label{sec1}
One of the most striking features of phase separation is the generation of long range correlations confined to the interfacial region. This fact has been first established within the framework of inhomogeneous fluids in a seminal paper by Wertheim \cite{Wertheim_1976}; we refer to \cite{RowlinsonWidom} for a historical account on these aspects and to \cite{Widom_1972, Evans_79, Jasnow_1984, gennes_wetting_1985, sullivantelodagama, dietrich_wetting_1988, MSchick, FLN, BEIMR} for reviews on interfaces and wetting phenomena. Descriptions based on the capillary wave model \cite{BLS_1965} have been proposed as effective frameworks for the characterization of correlations within the interfacial region \cite{Weeks_1977, BW_1985}. Further elaborations of these models have been developed in order to provide accurate descriptions of realistic systems \cite{MD, PRWE_2014} and have been flanked by accurate full scale numerical studies based on molecular dynamics simulations \cite{HD_2015}.

The manifestation of long range correlations at the interface separating coexisting bulk phases is generally investigated in momentum space through the notion of interface structure factor \cite{MD, BKV, PRWE_2014,CT_2016}. This tendency is actually triggered by the fact that results obtained from scattering experiments -- either with neutrons or X-rays -- probe correlations in momentum space; see e.g. \cite{HD_2015} and references therein. Although effective descriptions relying on the notion of interface -- such as capillary wave models and refinements thereof \cite{Weeks_1977, BW_1985} -- can be used in order to find correlation in real space, the first-principle derivation of the exact analytic form of these correlations from the underlying field theory has been obtained in \cite{DS_twopoint}. More recently, exact results for $n$-point correlation functions in two-dimensional systems exhibiting phase separation have been obtained in \cite{Squarcini_Multipoint}.

The two-dimensional case turns out to be very interesting because the scenario is dominated by strong thermal fluctuations and non-perturbative techniques can be used in order to find exact results. Among these findings, a central importance is played by those obtained by exploiting the exact solvability of the Ising model on the lattice with boundary conditions leading to phase separation \cite{Abraham_review}. In more recent times, it has been possible to formulate an exact field theory of phase separation \cite{DV} which encompasses a wide range of universality classes in two dimensions. The language of field theory proved to be successful in describing multifaceted aspects of interfacial phenomena in near-critical systems ranging from interface structure \cite{DV}, interfacial wetting transition \cite{DS_bubbles}, wetting transition on flat walls \cite{DS_wetting}, wedge willing transitions \cite{DS_wedge, DS_wedgebubble}, interface localization \cite{Delfino_localization}, interfacial correlations \cite{DS_twopoint,Squarcini_Multipoint}, and the interplay of geometry on correlations \cite{ST_droplet}.

The verification of theoretical predictions by means of Monte Carlo simulations is an invaluable test bench for the theory \cite{BLM}. The structure of single interfaces \cite{DV} and the occurrence of interfacial adsorption predicted in \cite{DS_bubbles} has been confirmed in \cite{DSS1}. We refer to \cite{DSS_2020,DSS_2021} for recently obtained analytical and numerical results about phase separation in three dimensions\footnote{See \cite{Delfino_vortex} for the extension of the theory to topological defect lines and to \cite{DSS_2019} for comparison with numerical simulations.}. This paper presents the comparison between theoretical and numerical results for interfacial correlations in the two dimensional Ising model. To be definite, we consider the near-critical regime of the two-dimensional Ising model at phase coexistence. The system is studied on the two-dimensional strip of width $R$ much larger than the bulk correlation length $\xi$. The observables considered in this paper are: magnetization and energy density profiles, two- and three-point correlation functions of the order parameter field and the passage probability of the off-critical interface.
For all the aforementioned quantities we provide closed-form analytic expressions which then we test through high-quality Monte Carlo (MC) simulations.

This outline of this paper is as follows. In Sec.~\ref{sec2}, we set up the calculation of the energy density profile across an interface and we also recall the main ideas involved in the probabilistic interpretation of fluctuating interfaces, which include the notions of passage probability and interface structure. We then compare the theoretical predictions for the energy density profile with MC simulations. Next, we consider the order parameter profile and its leading and first subleading finite-size corrections. At leading order, the order parameter profile is extracted from an exact probabilistic interpretation \cite{DV}. The first subleading correction, which occurs at order $\xi/R$, is borrowed from \cite{Squarcini_Multipoint}. Both the quantities are found to be in agreement with the numerical results. We conclude Sec.~\ref{sec2} by showing the comparison between theory and numerics for the passage probability, the latter is \emph{directly} extracted by sampling interface crossings on the lattice. Analytic expressions and numerical results for two- and three-point correlation functions of the order parameter field are presented in Secs.~\ref{sec3} and \ref{sec4}, respectively. Conclusive remarks are summarized in Sec.~\ref{sec5}. Appendix \ref{appendixB} summarizes various mathematical details involved in the calculation of two- and three-point correlation functions. Appendix \ref{AppendixC} shows how B\"urmann series can be fruitfully applied in order to characterize the asymptotic behavior of certain three-point correlation functions. Appendix \ref{AppendixD} collects results about mixed three-point correlation functions involving two order parameter fields and one energy density field.

\section{Theory of phase separation on the strip}
\label{sec2}
In this section, we review the exact field theoretical approach to phase separation and interfacial phenomena in two dimensions \cite{DV}. Our presentation follows closely the one outlined in \cite{DS_bubbles}, however, instead of presenting the field-theoretical calculation of the magnetization profile, we focus on the energy density profile. The reason for such a viewpoint relies on the fact that, as we are going to show, energy density correlation functions are proportional to the passage probability, a notion which completely characterizes the statistics of interfacial fluctuations. Anticipating some results, exact order parameter profiles and correlation functions involving the spin field will be computed within a probabilistic interpretation based on the passage probability extracted from energy density correlations \cite{DV, Squarcini_Multipoint}. Although several conclusions which we will drawn are valid for several universality classes in two dimensions, we will focus both the theory and the numerics to the Ising model.

As a warmup and in order to set the notations, we begin by recalling the lattice hamiltonian of the Ising model
\begin{equation}
\label{2001}
\mathcal{H} = - J \sum_{\langle i,j \rangle} s_{i}s_{j} \, ,
\end{equation}
with spins $s_{i}\in\{\pm1\}$ and the sum is restricted to nearest neighboring sites of a square lattice. The global $\mathbb{Z}_{2}$ symmetry corresponding to sign reversal of all spins is spontaneously broken below the critical temperature $T_{\textrm{c}}/J=2/\log(1+\sqrt{2})=2.691\,853 \dots$, in correspondence of which the model exhibits a second order phase transition \cite{Onsager_44}.

In this paper, we consider the two-dimensional Ising model along the phase coexistence line\footnote{The phase coexistence line is defined by the set of points in the phase diagram in which $H=0$ and $T<T_{\textrm{c}}$, where $H$ is the bulk magnetic field.} close enough to the critical temperature. The scaling limit of the lattice model in the closeness of $T_{\rm c}$ is described by a Euclidean field theory in the two-dimensional plane with coordinates $x$ and $y$. The aforementioned Euclidean field theory can be regarded as the analytic continuation to imaginary time $t= {\rm i} y$ of a relativistic field theory in a $(1+1)$-dimensional space time. Elementary excitations in $1+1$ dimensions are stable kink states $\vert K_{-+}(\theta) \rangle$ which interpolate between two different vacua, denoted $\vert \Omega_{-}\rangle$ and $\vert \Omega_{+} \rangle$, and analogously for $\vert K_{+-}(\theta) \rangle$. These topological excitations are relativistic particles with energy-momentum
\begin{equation}
\label{2002}
(e,p) = m \left(\cosh\theta,\sinh\theta\right) \, ,
\end{equation}
where $\theta$ is the rapidity and $m$ is the kink mass. In the Ising field theory there are only two degenerate vacua, the latter correspond respectively to pure phases in which the system is translationally invariant and ferromagnetically ordered. Pure phases can be selected by fixing the spins on a finite boundary and then by taking the thermodynamic limit in which the boundary is sent to infinity.

\begin{figure}[htbp]
\centering
\begin{tikzpicture}[thick, line cap=round, >=latex, scale=0.5]
\tikzset{fontscale/.style = {font=\relsize{#1}}}
\draw[thin, dashed, ->] (-10, 0) -- (11, 0) node[below] {$x$};
\draw[thin, dashed, -] (0, -5) -- (0, 5) node[left] {};
\draw[thin, dashed, ->] (0, 6.2) -- (0, 7) node[left] {$y$};
\draw[very thick, red, -] (0, 5) -- (10, 5) node[] {};
\draw[very thick, red, -] (0, -5) -- (10, -5) node[] {};
\draw[very thick, blue, -] (-10, 5) -- (0, 5) node[] {};
\draw[very thick, blue, -] (-10, -5) -- (0, -5) node[] {};
\draw[thin, fill=white] (0, -5) circle (3pt) node[below] {$(0,-R/2)$};;
\draw[thin, fill=white] (0, 5) circle (3pt) node[above] {$(0,R/2)$};;
\draw[thin, fill=white] (-5, 5) circle (0pt) node[above] {${\color{blue}{-}}$};;
\draw[thin, fill=white] (-5, -5) circle (0pt) node[below] {${\color{blue}{-}}$};;
\draw[thin, fill=white] (5, 5) circle (0pt) node[above] {${\color{red}{+}}$};;
\draw[thin, fill=white] (5, -5) circle (0pt) node[below] {${\color{red}{+}}$};;
\end{tikzpicture}
\caption{The strip geometry with $-+$ boundary conditions.}
\label{fig_geometry_1}
\end{figure}
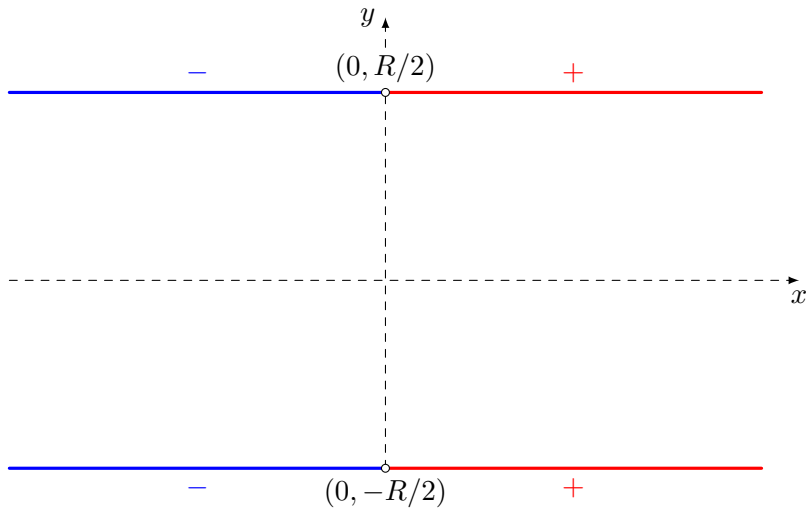

We study phase separation on a strip of width $R$ with fixed boundary conditions such that the boundary spins take the value $-1$ on the left side ($x<0$) and the value $+1$ on the right side ($x>0$); see Fig.~\ref{fig_geometry_1}. These boundary conditions lead to the emergence of phase separation when $R$ becomes much larger than the bulk correlation length $\xi$. The bulk correlation length describes the large-distance exponential decay of the connected spin-spin correlation function in pure phases \cite{Abraham78, McCoyWu, WMCTB}, i.e.\footnote{For large distances the exponential decay is multiplied by a power law which is not essential to recall here.},
\begin{equation}
\label{2003}
\langle \sigma(0,0) \sigma(x,y) \rangle_{\pm}^{\textrm{c}} \sim \textrm{e}^{-r/\xi} \, , \qquad r=\sqrt{x^{2}+y^{2}} \, .
\end{equation}
The kink mass $m$ turns out to be inversely proportional to the bulk correlation length; $\xi=1/(2m)$ in the low-temperature phase and $\xi=1/m$ in the high-temperature phase.

The switching of boundary condition from $-$ to $+$ at $x=x_{0}$ and time $t$ is implemented through the boundary state $\vert B_{-+}(x_{0};t) \rangle$, the latter can be decomposed over the complete basis of states of the bulk theory (the kinks states). Since the states entering in the aforementioned decomposition have to interpolate between the phases $-$ and $+$, we have
\begin{equation}
\label{2004}
\vert B_{-+}(x_{0};t) \rangle = \textrm{e}^{-\textrm{i}tH+\textrm{i}x_{0}P} \biggl[ \int_{\mathbb{R}}\frac{\textrm{d}\theta}{2\pi} f_{-+}(\theta) \vert K_{-+}(\theta) \rangle + \dots \biggr] \, ,
\end{equation}
where $H$ and $P$ are the energy and momentum operators of the $(1+1)$-dimensional quantum field theory. The ellipses correspond to multi-kink states with total mass larger than $m$. The lightest term appearing in the ellipses corresponds to the emission of a three-kink excitation $\vert K_{-+}(\theta_{1})K_{+-}(\theta_{2})K_{-+}(\theta_{3})\rangle$ which interpolates between the two vacua. In general, an arbitrary multi-kink state compatible with the topological charge imposed by the boundary must involve an odd number of kinks.

Within the one-kink approximation, which suffices in order to describe phase separation for large $R/\xi$, the partition function for the strip with $-+$ boundary conditions of Fig.~\ref{fig_geometry_1} reads
\begin{equation}
\begin{aligned}
\label{2005}
\mathcal{Z}_{-+}(R) & = \langle B_{-+}(x_{0};\textrm{i}R/2) \vert B_{+-}(x_{0};-\textrm{i}R/2) \rangle \, ,\\
& \simeq \int_{\mathbb{R}} \frac{\textrm{d}\theta}{2\pi} |f_{-+}(\theta)|^{2} \textrm{e}^{-mR\cosh\theta} \, .
\end{aligned}
\end{equation}
The symbol $\simeq$ stands for the omission of subleading terms stemming from heavier states. The normalization of kinks $\langle K_{ab}(\theta) \vert K_{a^{\prime}b^{\prime}}(\theta^{\prime}) \rangle=2\pi \delta_{aa^{\prime}}\delta_{bb^{\prime}}\delta(\theta-\theta^{\prime})$ has been used in (\ref{2005}). The limit of large $R/\xi$ we are interested in amounts to project the integrand at small rapidities; hence, a standard saddle-point calculation yields
\begin{equation}
\label{2006}
\mathcal{Z}_{-+}(R) \simeq |f_{-+}(0)|^{2} \frac{\textrm{e}^{-mR}}{\sqrt{2\pi mR}} \, .
\end{equation}

The occurrence of phase separation can be detected by a local measurement of the spin field, the latter amounts to compute the order parameter profile $\langle \sigma(x,y) \rangle_{-+}$. The magnetization profile $\langle \sigma(x,y) \rangle_{-+}$ interpolates between the asymptotic values $-M$ and $+M$ where $M$ is the spontaneous magnetization of bulk phases in the far left and right regions, respectively. The jump of order parameter across the interface is accompanied by an increase of the energy density, which we are going to compute. The energy density for the Ising model on the lattice can be defined by $\varepsilon_{i} = -\sum_{j \sim i}\sigma_{i}\sigma_{j}$, where the sum runs over nearest neighbors of site $i$ and the overall factor $-1$ is purely conventional. Contrary to the order parameter -- which corresponds to an extended spin field configuration -- the energy density profile is localized in the sense that it exhibits a nontrivial dependence through the coordinates only within the interfacial region, while away from the interface it attains the bulk value $\langle\varepsilon\rangle$ in both phases.

The energy density profile is defined as follows
\begin{equation}
\label{2009}
\langle \varepsilon(x,y) \rangle_{-+} = \frac{ \langle B_{-+}(0;\textrm{i}R/2) \vert \varepsilon(x,y) \vert B_{+-}(0;-\textrm{i}R/2) \rangle }{ \langle B_{-+}(0;\textrm{i}R/2) \vert B_{+-}(0;-\textrm{i}R/2) \rangle } \, .
\end{equation}
The field entering in (\ref{2009}) can be translated to the origin thanks to
\begin{equation}
\label{2010}
\varepsilon(x,y) = \textrm{e}^{\textrm{i}xP+yH} \varepsilon(0,0) \textrm{e}^{-\textrm{i}xP-yH} \, ,
\end{equation}
by using the boundary state (\ref{2004}) the one-point correlation function (\ref{2009}) reads
\begin{equation}
\begin{aligned}
\label{2011}
\langle \varepsilon(x,y) \rangle_{-+} & = \frac{1}{\mathcal{Z}_{-+}(R)}\int_{\mathbb{R}^{2}} \frac{\textrm{d}\theta_{1}\textrm{d}\theta_{2}}{(2\pi)^{2}} f_{-+}^{*}(\theta_{1}) f_{-+}(\theta_{2}) \times \\
& \times \langle K_{-+}(\theta_{1}) \vert \textrm{e}^{(-R/2+y)H} \varepsilon(x,0) \textrm{e}^{-(R/2+y)H} \vert K_{-+}(\theta_{2}) \rangle + \dots \, ,
\end{aligned}
\end{equation}
with $|y|<R/2$. In analogy with (\ref{2004}), the ellipses denote terms coming from multi-kink states whose contribution is subleading with respect to the term shown in (\ref{2011}). Thus, we have
\begin{equation}
\begin{aligned}
\label{2012}
\langle \varepsilon(x,y) \rangle_{-+} & \simeq \frac{1}{\mathcal{Z}_{-+}(R)}\int_{\mathbb{R}^{2}} \frac{\textrm{d}\theta_{1}\textrm{d}\theta_{2}}{(2\pi)^{2}} f_{-+}^{*}(\theta_{1}) f_{-+}(\theta_{2}) \mathcal{M}_{-+}^{\varepsilon}(\theta_{1}\vert\theta_{2}) O(\theta_{1},\theta_{2}) \, ,
\end{aligned}
\end{equation}
where $O(\theta_{1},\theta_{2}) = \exp\bigl[ - M_{-} \cosh\theta_{1} - M_{+} \cosh\theta_{2} + \textrm{i} mx (\sinh\theta_{1}-\sinh\theta_{2})\bigr]$, $M_{\pm}=m(R/2\pm y)$, and
\begin{equation}
\label{2013}
\mathcal{M}_{-+}^{\varepsilon}(\theta_{1}\vert\theta_{2}) = \langle K_{-+}(\theta_{1}) \vert \varepsilon(0,0) \vert K_{-+}(\theta_{2}) \rangle \, .
\end{equation}
The matrix element (\ref{2013}) can be decomposed into a connected part and a disconnected one; thus,
\begin{equation}
\label{2014}
\mathcal{M}_{-+}^{\varepsilon}(\theta_{1}\vert\theta_{2}) = F_{\varepsilon}(\theta_{12}+\textrm{i}\pi) + 2\pi \langle \varepsilon \rangle\delta(\theta_{12}) \, ,
\end{equation}
where $\theta_{12}\equiv\theta_{1}-\theta_{2}$, and $F_{\varepsilon}(\theta_{12}+\textrm{i}\pi)$ is the two-particle form factor of the energy density field \cite{Delfino_review}. Since the large $mR$ asymptotic behavior projects the integrand at small rapidities, the leading asymptotic behavior of the integral is encoded in the infrared (low-energy) properties of the bulk and boundary form factors, respectively $F_{\varepsilon}(\theta_{12}+\textrm{i}\pi)$ and $f_{-+}(\theta)$.

By virtue of reflection symmetry the boundary amplitude satisfies $f_{-+}(\theta)=f_{+-}(-\theta)$. Moreover, since the phases $-$ and $+$ play a symmetric role, there is invariance under exchange of labels, i.e., $f_{-+}(\theta)=f_{+-}(\theta)$. These observation imply the low-rapidity behavior $f_{-+}(\theta)=f_{0}+f_{2}\theta^{2}+O(\theta^{4})$ with $f_{0} \equiv f_{-+}(0)$ \cite{DV}. We recall that the energy density field is proportional to the trace of the stress tensor field $\Theta(x,y)$, i.e., $\Theta(x,y) \propto m \varepsilon(x,y)$. Furthermore, the form factor of the stress tensor satisfies the normalization $F_{\Theta}(\textrm{i}\pi)=2\pi m^{2}$ \cite{MS} and, without loss of generality, we can set the normalization of the energy density form factor to be
\begin{equation}
\label{2015}
F_{\varepsilon}(\textrm{i}\pi) = \mathcal{C}_{\varepsilon} m \, ,
\end{equation}
where $\mathcal{C}_{\varepsilon}$ is a proportionality constant which depends on the specific normalization of the energy density field and its implementation on the lattice. By inserting (\ref{2014}) into (\ref{2012}) the leading-order term in the low-rapidity expansion yields
\begin{equation}
\begin{aligned}
\label{2016}
\langle \varepsilon(x,y) \rangle_{-+} & \simeq \langle \varepsilon \rangle + \frac{F_{\varepsilon}(\textrm{i}\pi)|f_{0}|^{2}\textrm{e}^{-mR}}{\mathcal{Z}_{-+}(R)}  \int_{\mathbb{R}^{2}} \frac{\textrm{d}\theta_{1}\textrm{d}\theta_{2}}{(2\pi)^{2}} \textrm{e}^{-M_{-}\theta_{1}^{2}/2-M_{+}\theta_{2}^{2}/2+\textrm{i}mx\theta_{12}} \, ,
\end{aligned}
\end{equation}
the symbol $\simeq$ stands for the omission of terms at order $O(R^{-3/2})$, which are thus subleading with respect to the $O(R^{-1/2})$ term displayed in (\ref{2016}). The calculation of the factorized Gaussian integrals appearing in (\ref{2016}) is immediate. The connected correlation function
\begin{equation}
\label{2017}
\mathcal{G}_{\varepsilon}^{\textrm{c}}(x,y) = \langle \varepsilon(x,y) \rangle_{-+}- \langle \varepsilon \rangle
\end{equation}
reads
\begin{equation}
\begin{aligned}
\label{2018}
\mathcal{G}_{\varepsilon}^{\textrm{c}}(x,y) & = \mathcal{C}_{\varepsilon} \frac{\textrm{e}^{-\chi^{2}}}{\sqrt{\pi}\kappa\lambda} \, ;
\end{aligned}
\end{equation}
the dependence through the coordinates $x$ and $y$ is encoded in the variables $\chi$ and $\kappa$ defined by
\begin{equation}
\label{2019}
\chi = \frac{\eta}{\kappa} \, , \qquad \kappa=\sqrt{1-\tau^{2}} \, , \qquad \tau=2y/R \, ,
\end{equation}
in the above, $\eta=x/\lambda$ is the rescaled horizontal coordinate and $\lambda=\sqrt{R/(2m)}=\sqrt{R\xi}$.

Far away from the interfacial region, i.e., $|x|\gg\lambda$, the energy density profile $\langle \varepsilon(x,y) \rangle_{-+}$ approaches the bulk value $\langle \varepsilon \rangle$. On the other hand, in the closeness of the interfacial region, i.e., $|x|\ll \lambda$, the energy density exhibits a deviation from the bulk value. The deviation is due to local increase of the disorder in the region where the two coexisting bulk phases come in touch. The deviation along the $x$-axis reaches its maximum at $x=0$ and is given by
\begin{equation}
\label{2020}
\mathcal{G}_{\varepsilon}^{\textrm{c}}(0,y) = \frac{\mathcal{C}_{\varepsilon}}{\sqrt{\pi R \xi} \sqrt{1-(2y/R)^{2}}} \, .
\end{equation}
We notice that (\ref{2020}) diverges upon approaching the pinning points $(0,\pm R/2)$. It is worth observing that the total excess energy on the strip is finite and it is given by
\begin{equation}
\label{2021}
\int_{\mathbb{R}}\textrm{d}x \int_{-R/2}^{R/2}\textrm{d}y \, \mathcal{G}_{\varepsilon}^{\textrm{c}}(x,y) = \mathcal{C}_{\varepsilon}R \, .
\end{equation}
We stress that (\ref{2021}) is valid in the near-critical region where the field-theoretical formalism applies. The right hand side of (\ref{2021}) is positive due to our normalization of the energy density field; see Sec.~\ref{sec_nr}. Rigorous results obtained from low-temperature expansions show that the integrated energy density correlation function is proportional to $-\textrm{d}\Sigma/\textrm{d}\beta$, where $\beta=1/T$ and $\Sigma$ is the surface tension of the $ab$ interface \cite{Bricmont}. It is straightforward to realize that the quantity $-\textrm{d}\Sigma/\textrm{d}\beta$ is positive in the closeness of the critical temperature. This can be realized by recalling that $\Sigma$ is related to the kink mass via $\Sigma = m$ \cite{DV} and in the near-critical region $\Sigma \sim (T_{\rm c}-T)^{\mu}$ (clearly, with a positive pre-factor). Moreover, $m \propto 1/\xi$ and $\xi \sim (T_{\rm c} - T)^{-\nu}$, with $\nu=1$ for the Ising model. By combining the above, we find $\mu=1$, which is compatible with Widom's hyperscaling relation $\mu=(d-1)\nu$ \cite{Widom_1972} in $d=2$. Of course, $\mu=1$ emerges also from the exact expression for the surface tension \cite{Abraham_review}.

\subsection{Numerical results}
\label{sec_nr}
We can now present the comparison between theory and numerics. We have performed Monte Carlo simulations on a finite rectangle with horizontal length $L \gtrsim R$ and temperature $T$ such that $\xi \ll R$. Without loss of generality, we set $J=1$ in simulations. Our hybrid Monte Carlo scheme (see, e.g. \cite{LandauBinder}) combines the standard Metropolis algorithm and the Wolff cluster algorithm \cite{Wolff}. The minimum number of MC steps per site is $10^{7}$. Parallelization was obtained by independently and simultaneously simulating up to 128 Ising lattices on a parallel computer. An appropriately seeded family of dedicated, very large period, Mersenne Twister random number generators \cite{MN_1998b}, in the MT2203 implementation of the Intel Math Kernel Library, was used in order to simultaneously generate independent sequences of random number to be used for the MC updates of the lattices.

\begin{figure}[htbp]
\centering
\includegraphics[width=105mm]{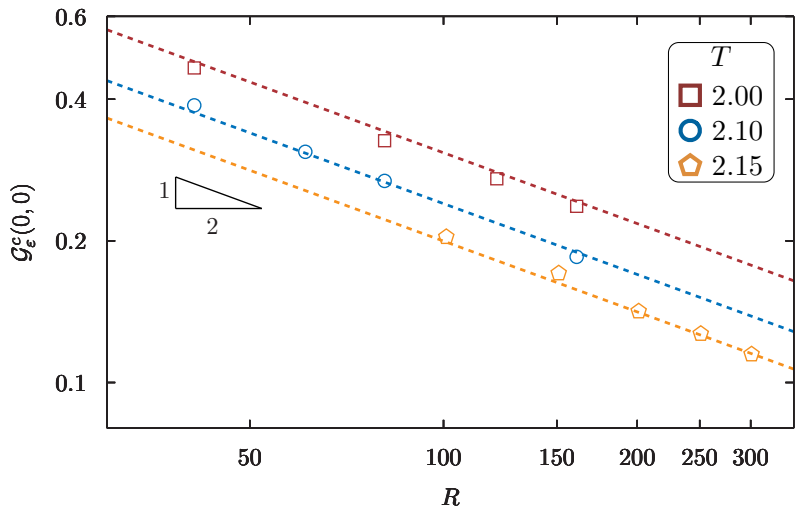}
\caption{Maximum of the energy density as a function of $R$ in log-log plot. Symbols indicate different temperatures: $T=2.00$ ($\square$), $T=2.10$ ($\varbigcirc$), $T=2.15$ ($\pentagon$). Straight dashed lines correspond to the scaling behavior $\propto R^{-1/2}$ predicted by (\ref{2020}).}
\label{fig_epsmax}
\end{figure}

The theory predicts that the maximum of the energy density $\mathcal{G}_{\varepsilon}^{\textrm{c}}(0,0)$ (with respect to $x$) scales as $R^{-1/2}$ at fixed temperature, while for fixed $R$ the maximum of the energy density depends on the temperature through $\xi^{-1/2}$. The above scaling behaviors are confirmed by the numerical results in Fig.~\ref{fig_epsmax}. Straight lines in the log-log plot correspond to the scaling with $R^{-1/2}$.
The temperature dependence shown in Fig.~\ref{fig_epsmax} is due to the bulk correlation length via the factor $\xi^{-1/2}$. For the planar Ising model in the low-temperature phase
\begin{equation}
\label{2022}
\xi = (4K-4K^{\star})^{-1} \, ,
\end{equation}
with the dual coupling $K^{\star}$ defined by means of $\exp(-2K^{\star})=\tanh K$ with $K=J/T$ \cite{Abraham_review}. From the numerical simulations, we extract the overall (non universal) amplitude $\mathcal{C}_{\varepsilon}=4.53$. Since $\mathcal{C}_{\varepsilon}$ is positive, the energy density increases upon approaching the interfacial region. This expected feature indicates a local increase of disorder with respect to the bulk phases, as we anticipated.

\begin{figure}[htbp]
\centering
\includegraphics[width=105mm]{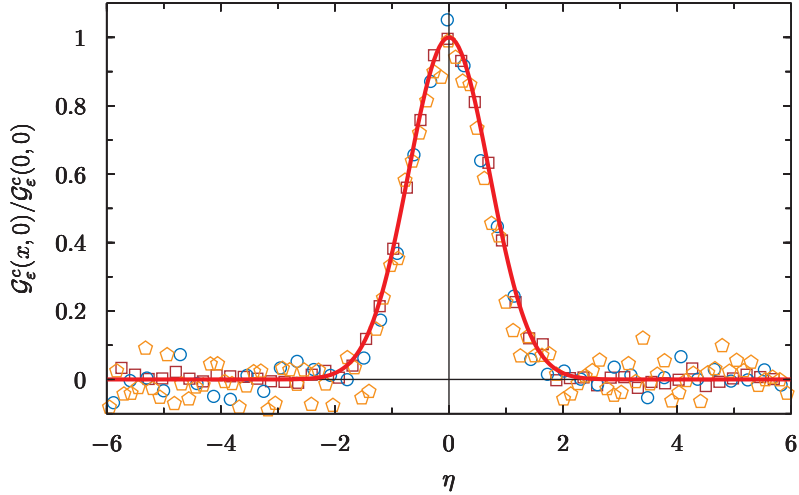}
\caption{Scaling function for the energy density profile. Data sets: $T=2.00$, $R=201$, $L=352$ ($\square$), $T=2.10$, $R=81$, $L=252$ ($\varbigcirc$), $T=2.15$, $R=301$, $L=602$ ($\pentagon$).}
\label{fig_eps}
\end{figure}

In Fig.~\ref{fig_eps}, we test the theoretical prediction (\ref{2018}) for the spatial dependence of the connected energy density profile. From the profiles $\langle \varepsilon(x,0) \rangle_{-+}$ obtained within numerical simulations, we read the offset value $\langle \varepsilon \rangle$, which is the quantity we subtract in (\ref{2017}). We observe that $\langle \varepsilon \rangle$ obtained in our simulations perfectly agrees with the theoretical value known in the literature\footnote{Notice that our definition of the energy density differs from the one given in \cite{McCoyWu} due to a different convention relative to the summation of neighboring sites; in our case the sum runs over the coordination number of the square lattice, which is $4$.} \cite{McCoyWu}.
Thanks to a rescaling of the horizontal coordinate, via $\eta=x/\lambda$, numerical results at different temperatures and lattice sizes collapse onto a single scaling curve given by $\mathcal{G}_{\varepsilon}^{\textrm{c}}(x,0)/\mathcal{G}_{\varepsilon}^{\textrm{c}}(0,0)=\textrm{e}^{-\eta^{2}}$, which is the continuous curve plotted in Fig.~\ref{fig_eps}.

\subsection{Probabilistic interpretation}
\label{sec2.2}
We can reconstruct the energy density profile by adapting the probabilistic approach described in \cite{DV}. Regarding the interface as a sharp line separating the left and right phases, we stipulate that it crosses the interval $(x,x+\textrm{d}x)$ at ordinate $y$ with probability $P_{1}(x,y)\textrm{d}x$. The expectation value of the energy density field can be obtained by weighting the energy density profile
\begin{equation}
\label{2025}
\varepsilon(x\vert u) = \langle\varepsilon\rangle + A_{\varepsilon}^{(0)} \, \delta(x-u) + \dots
\end{equation}
with the passage probability $P_{1}(x,y)$. The energy density profile $\varepsilon(x\vert u)$ gives the energy density at the point $(x,y)$ when the interface crosses the interval $(u,u+\textrm{d}u)$ at ordinate $y$. Since the energy density takes the same value in both phases, the expansion (\ref{2025}) starts with the bulk expectation value $\langle\varepsilon\rangle$. As a result, the sum over interfacial configurations reads
\begin{equation}
\label{2024}
\langle \varepsilon(x,y) \rangle_{-+} = \int_{\mathbb{R}}\textrm{d}u \, P_{1}(u,y) \varepsilon(x\vert u) \, .
\end{equation}
By matching the field-theoretical calculation (\ref{2017}) with the averaging procedure implied by (\ref{2024}), we extract the passage probability
\begin{equation}
\label{2026}
P_{1}(x,y) = \frac{\textrm{e}^{-\chi^{2}}}{\sqrt{\pi}\kappa\lambda} \, ,
\end{equation}
and the structure amplitude
\begin{equation}
\label{2027}
A_{\varepsilon}^{(0)} = \frac{F_{\varepsilon}(\textrm{i}\pi)}{m} = \mathcal{C}_{\varepsilon} \, .
\end{equation}
Subsequent corrections can be determined in a systematic fashion by taking into account further terms in the low-energy expansion of bulk and boundary form factors in the field-theoretic calculation which lead us to the leading order result (\ref{2017}). Being $P_{1}(x,y)$ a probability density, it is normalized as follows $\int_{\mathbb{R}}\textrm{d}x \, P_{1}(x,y)=1$.

The probability density (\ref{2026}) is the one of a Brownian bridge in which the time is identified with the $y$ coordinate. The endpoints $x=0$, $y=\pm R/2$ of the Brownian bridge correspond to the points in which the interface is pinned on the boundaries. In particular, (\ref{2026}) implies that midpoint fluctuations of the interface grow as the square root of the separation between pinning points; such an observation has been rigorously proved for low temperatures \cite{Gallavotti_72}. The convergence of interface fluctuations towards the Brownian bridge has been proved for the Ising model \cite{GI} and for the $q$-state Potts model \cite{CIV}. However, field theory implies that the occurrence of Brownian bridges is a more general feature which emerges naturally in a larger variety of universality classes \cite{DV,DS_bubbles}.

Once we have extracted the passage probability, the probabilistic formulation allows us to reconstruct the magnetization profile by following the same guidelines which lead us to the energy density profile. Thus, the the magnetization profile is given by
\begin{equation}
\label{2030}
\langle \sigma(x,y) \rangle_{-+} = \int_{-\infty}^{+\infty}\textrm{d}u \, P_{1}(u;y) \sigma_{-+}(x \vert u)
\end{equation}
with the sharp magnetization profile
\begin{equation}
\label{2031}
\sigma_{-+}(x \vert u) = -M\theta(u-x)+M\theta(x-u) + \dots
\end{equation}
and $\theta(x)$ is Heaviside step function \cite{DV}.

The calculation of the magnetization profile within the field-theoretical approach can be refined in order to take into account the first subleading correction in a large-$R$ expansion. Focusing on the midline $y=0$, the result for the profile reads \cite{Squarcini_Multipoint}
\begin{equation}
\label{2034}
\langle \sigma(x,0) \rangle_{-+}/M = \textrm{erf}(\eta) + \frac{b\mathcal{B}(\eta)}{mR} + O(R^{-2}) \, ,
\end{equation}
where $\textrm{erf}(\cdots)$ is the error function \cite{Temme} and $M$ is the spontaneous magnetization, which for the two-dimensional Ising model on the square lattice is given by \cite{McCoyWu,Yang}
\begin{equation}
\label{2032}
M = \left( 1-(\sinh(2K))^{-4} \right)^{1/8} \, ,
\end{equation}
where $K=1/T$ (since $J=1$) and $T$ is the temperature. The subleading correction at order $R^{-1}$ occurs via the scaling function for the branching profile $\mathcal{B}(\eta)= \pi^{-1/2}\eta \, \textrm{e}^{-\eta^{2}}$. The overall amplitude is $b=(2/3) + 4f_{2}$. From the low-rapidity expansion of the boundary form factor \cite{LS}, we find $f_{2}=3/8$ and $b=13/6$.

\begin{figure}[htbp]
\centering
\includegraphics[width=105mm]{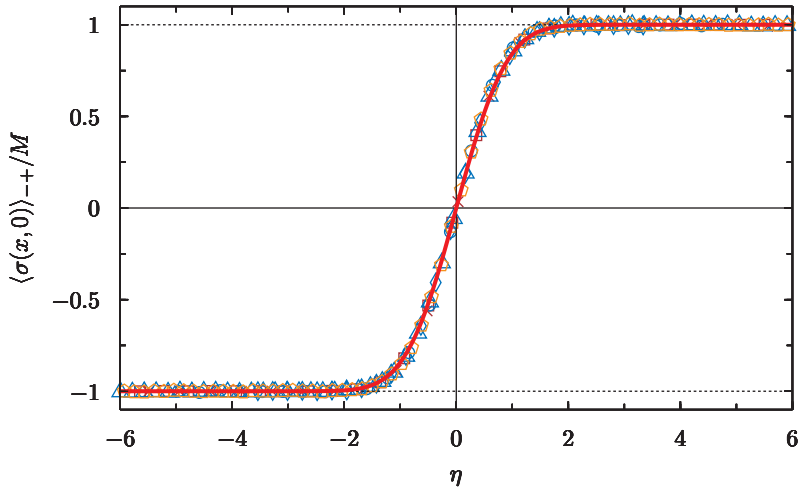}
\caption{Scaling function for the magnetization profile. Symbols correspond to the data sets $T=2.00$, $R=121$ ($\times$), $T=2.00$, $R=201$ ($\square$), $T=2.10$, $R=161$ ($\varbigcirc$), $T=2.10$, $R=301$ ($\triangle$), $T=2.10$, $R=321$ ($\diamond$), $T=2.15$, $R=301$ ($\pentagon$). The  solid red curve corresponds to the scaling function $\textrm{erf}(\eta)$.}
\label{fig_mag}
\end{figure}

\begin{figure}[htbp]
\centering
\includegraphics[width=105mm]{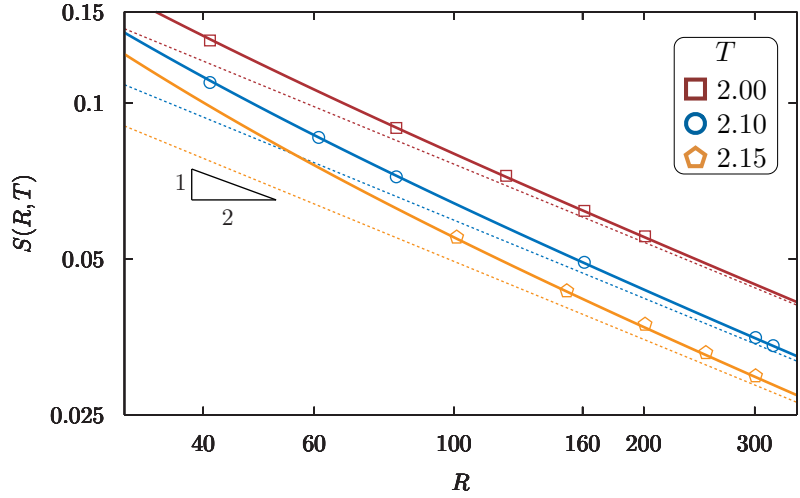}
\caption{Slope $S(R,T)$ as function of $R$ for various $T$. Results extracted from numerical simulations are illustrated with symbols. Dashed lines indicate the leading-order behavior $\propto R^{-1/2}$ given by (\ref{26022021_1444}) with $b=0$. Solid curves correspond to the result (\ref{26022021_1444}) including the subleading correction with $b=13/6$.}
\label{fig_slope}
\end{figure}

In Fig.~\ref{fig_mag}, we compare the numerical results obtained within MC simulations with the analytic result (\ref{2034}) at leading order. Data sets obtained at different temperatures $T$ and lattice width $R$ collapse onto the scaling function given in (\ref{2034}) with remarkably good accuracy, and without free parameters.
In order to test the expression (\ref{2034}) for the magnetization profile, we extract the slope of the profile at $x=0$ and compare it against the theoretical prediction implied by (\ref{2034}). The slope in the origin is given by
\begin{equation}
\begin{aligned}
\label{26022021_1444}
\partial_{x} \langle \sigma(x,0) \rangle_{-+}/M \big\vert_{x=0} & = \frac{2}{\sqrt{\pi R \xi}} \left( 1 + \frac{b\xi}{R} \right) + O(R^{-5/2}) \\
& \equiv S(R,T) \, .
\end{aligned}
\end{equation}
The theoretical result (\ref{26022021_1444}) is compared with the numerical data in Fig.~\ref{fig_slope}. The power-law behavior $\propto R^{-1/2}$ for large $R$ is visible in the log-log plot of Fig.~\ref{fig_slope}. We also point out that the corrective term at order $R^{-3/2}$ which appears in the bracket of (\ref{26022021_1444}) is crucial in order to establish a quantitative agreement between theory and numerics. As a further check, we have also fitted the numerical data for the slope with the right hand side of (\ref{26022021_1444}) with an unknown $b$ and found the optimal value $b=2.17$, which is remarkably close to the theoretical one $b=13/6=2.1\bar{6}$.

\begin{figure}[htbp]
\centering
\includegraphics[width=105mm]{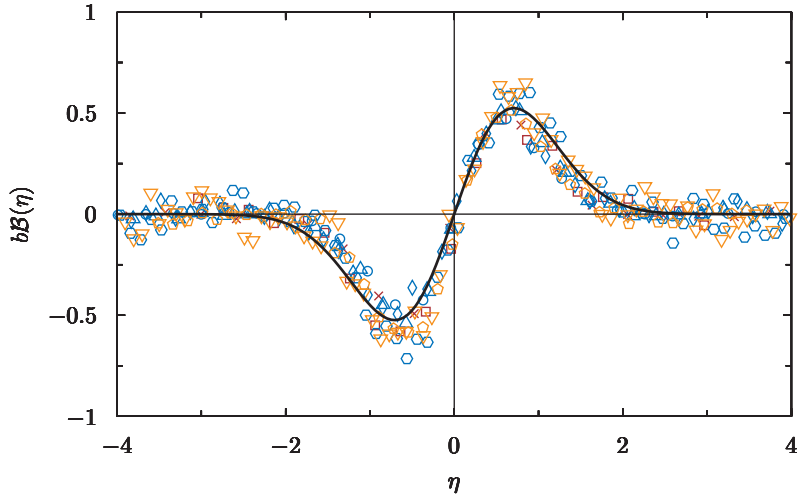}
\caption{Subleading correction of the magnetization profile. Numerical data:
$T=2.0$, $R=41$ ($\times$), 
$T=2.0$, $R=81$ ($\square$), 
$T=2.1$, $R=41$ ($\circ$), 
$T=2.1$, $R=61$ ($\triangle$), 
$T=2.1$, $R=81$ ($\diamond$), 
$T=2.1$, $R=301$ ($\hexagon$), 
$T=2.15$, $R=101$ ($\pentagon$), 
$T=2.15$, $R=301$ ($\triangledown$).
The theoretical result is indicated with the solid black line.}
\label{fig_branching}
\end{figure}

We can now push the comparison of the subleading profile $\mathcal{B}(\eta)$ for $x \neq 0$. To this end, we subtract from the numerical data the leading order form of the magnetization profile given by $\textrm{erf}(\eta)$ in (\ref{2034}), the result is then multiplied by $mR$. The numerical result obtained within this procedure is then compared against the theoretical prediction, which is the profile $b\mathcal{B}(\eta)$. The subtraction of the leading-order profile and the multiplication by the factor $mR$ increases the statistical noise, as one can clearly see by inspection of Fig.~ \ref{fig_branching}. Nonetheless, a good data collapse is observed for several values of $T$ and $R$.

Even in this case we can regard $b$ as unknown and seek for the best fit of the numerical data with the theoretical prediction of the subleading profile. We have tested several data sets obtained at temperatures $T=2.0$, $T=2.1$, and $T=2.15$ with the values of $R$ summarized in Fig.~\ref{fig_branching}. The fit obtained from the data sets of Fig.~\ref{fig_branching} yields $b=2.169$, in agreement with the analysis of the slope in $x=0$. Although statistical errors are visible, the quantitative agreement between theory and simulations is satisfactory

\subsection{Interface tracing on the lattice}
The remarkable agreement between theoretical and numerical results for spin and energy density profiles provides an indirect validation of the probabilistic interpretation. Although the Brownian bridge property is known from rigorous results, finding passage probabilities in a \emph{direct} fashion is an interesting problem on its own, especially for those circumstances in which exact results are not yet available. We will provide further remarks on this point in the concluding section. The idea of extracting passage probabilities for extended objects has been already successfully employed in the study of critical interfaces by means of the Schramm-Loewner Evolution (SLE), in investigations of critical spin clusters, and geometrical properties of percolative observables; see \cite{HK, Cardy_SLE, BB2006} and references therein. Since we are interested in the strictly subcritical regime, we will provide a \emph{direct} measurement of the passage probability for \emph{off-critical} interfaces by means of numerical simulations. The recipe we are going to discuss applies to the critical case too.

\begin{figure*}[htbp]
\centering
        \begin{subfigure}[b]{0.45\textwidth}
            \centering
            \includegraphics[width=\textwidth]{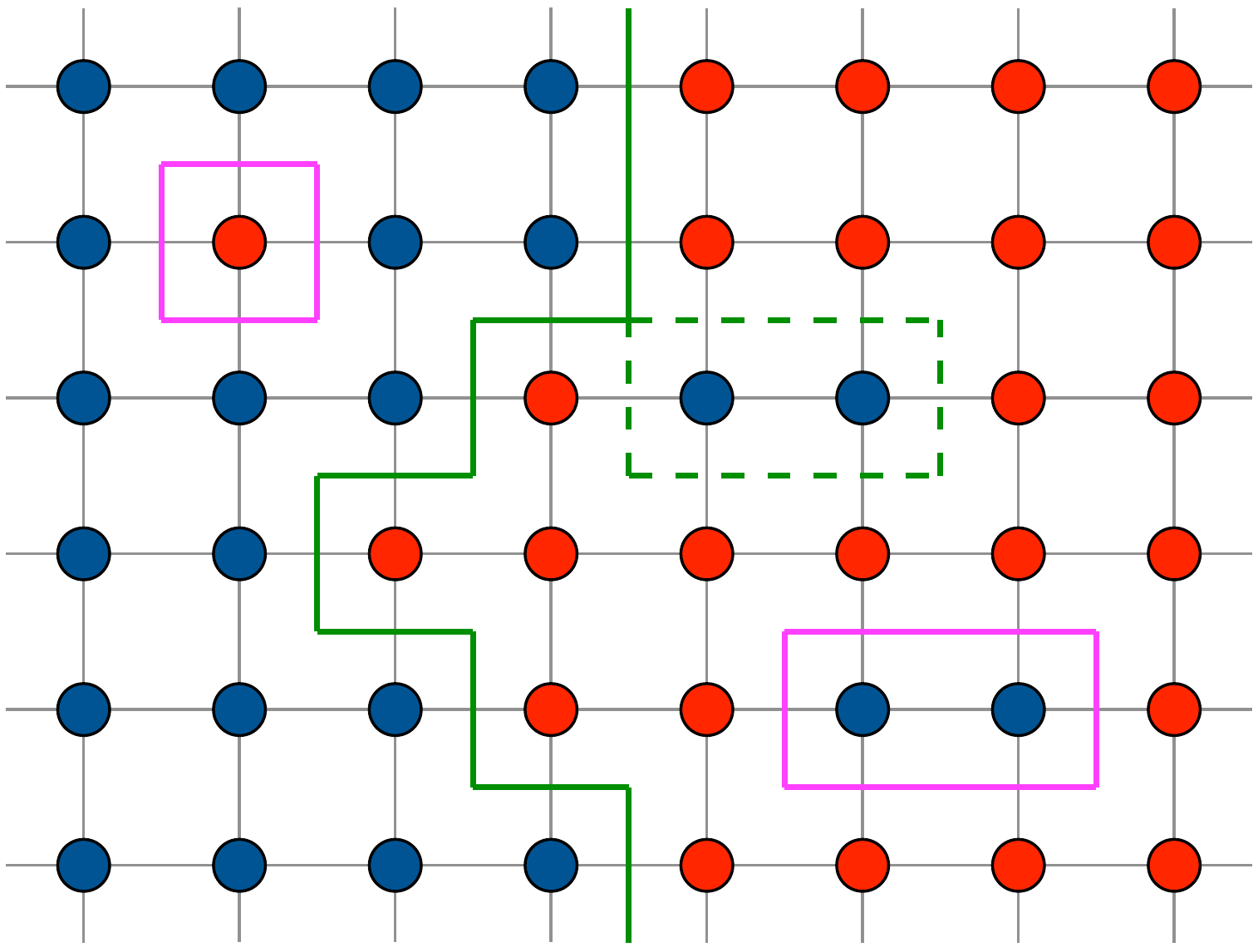}
            \caption[]%
            {{\small Interface on the dual lattice.}}    
            \label{fig_real_medial_a}
        \end{subfigure}
\hfill
        \begin{subfigure}[b]{0.45\textwidth}  
            \centering 
            \includegraphics[width=\textwidth]{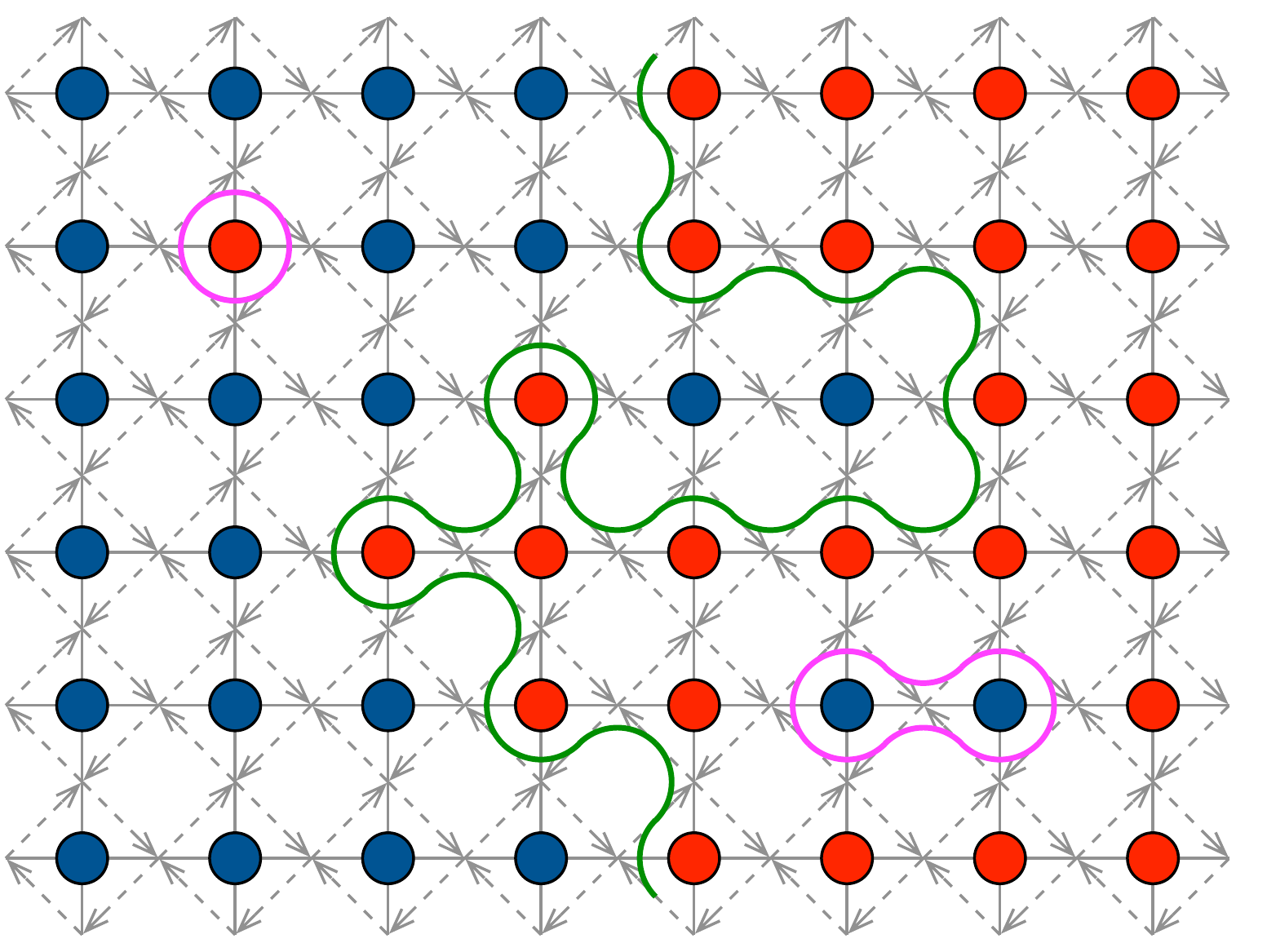}
            \caption[]%
            {{\small Interface on the medial lattice.}}    
            \label{fig_real_medial_b}
        \end{subfigure}
\vskip\baselineskip
\caption[]
{\small Construction of the interface on the square lattice (a) and on the medial lattice (b).}
\label{fig_real_medial}
\end{figure*}

\begin{figure*}[htbp]
\centering
\hspace{10mm}
        \begin{subfigure}[b]{0.15\textwidth}
            \centering
            \includegraphics[width=\textwidth]{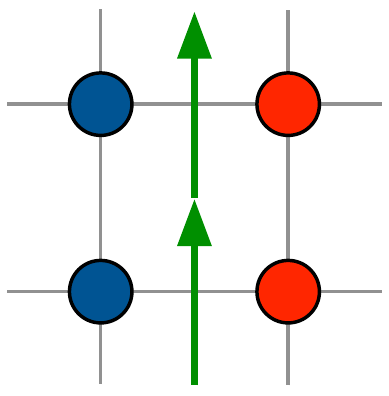}
            \caption[]%
            {{\small }}    
            \label{fig_plaquette_a}
        \end{subfigure}
\hfill
        \begin{subfigure}[b]{0.15\textwidth}
            \centering
            \includegraphics[width=\textwidth]{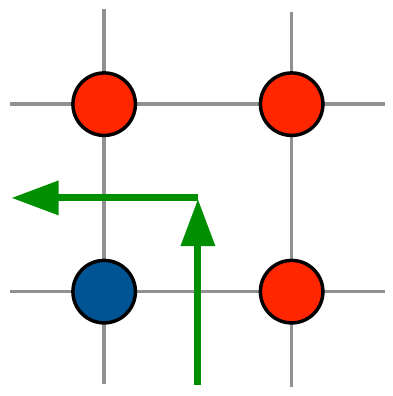}
            \caption[]%
            {{\small }}    
            \label{fig_plaquette_b}
        \end{subfigure}
\hfill
        \begin{subfigure}[b]{0.15\textwidth}
            \centering
            \includegraphics[width=\textwidth]{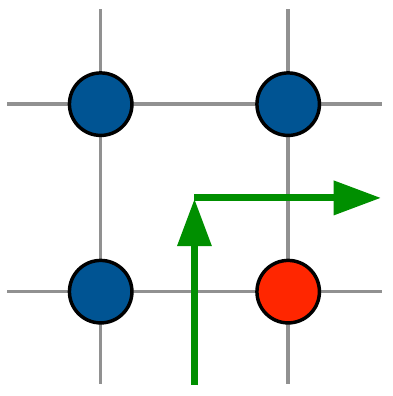}
            \caption[]%
            {{\small }}    
            \label{fig_plaquette_c}
        \end{subfigure}
\hfill
        \begin{subfigure}[b]{0.15\textwidth}
            \centering
            \includegraphics[width=\textwidth]{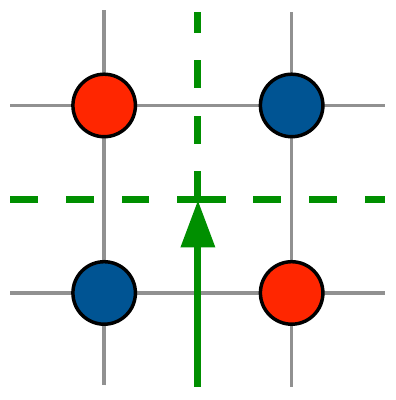}
            \caption[]%
            {{\small }}    
            \label{fig_plaquette_d}
        \end{subfigure}
\hspace{10mm}
\vskip\baselineskip
\caption[]
{\small Construction of the interface on square plaquettes. In (a)-(d) the pair of spins on the bottom defines a vertically oriented interface segment on the dual lattice. The interface then proceeds vertically in (a), turns left in (b), and turns right in (c). Plaquette (d) leads to an ambiguity in the definition of the interface.}
\label{fig_plaquette}
\end{figure*}

The line of separation between two coexisting phases in the Ising model is a well defined observable on the lattice \cite{Gallavotti}. For the square lattice, which is the one we are using in our simulations, the interface is constructed on the dual lattice by those dual bonds which cross real bonds connecting opposite spins; see Fig.~ \ref{fig_real_medial_a}. We can also regard the interface as the result of an exploration process originated in the lower pinning point $(x=0,y=-R/2)$ and propagated towards the upper pinning point $(x=0,y=+R/2)$.
Within this construction, the interface proceeds straight in vertical direction, turn left or turn right, as illustrated in the elementary plaquettes of Fig.~\ref{fig_plaquette_a}-\ref{fig_plaquette_c}, respectively. 
On the square lattice, however, the occurrence of the plaquette of Fig.~\ref{fig_plaquette_d} does not lead to a precise definition of the interface. One can certainly prescribe a way to resolve the ambiguity, e.g., by going straight. The standard way to circumvent the ambiguity is to pass on the medial lattice \cite{HK}. Interfaces on the medial lattice are constructed as follows: we draw a square lattice rotated by $45$ degrees with respect to the original one and assign a clockwise pattern of arrows to the medial bonds surrouding real lattice sites, the interface segments are then drawn by following the arrows with the prescription that the interface does not cross real bonds connecting identical spins; see Fig.~\ref{fig_real_medial_b}. By construction, interfaces on the medial lattice are always unambiguously defined.

\begin{figure}[htbp]
\centering
\includegraphics[width=105mm]{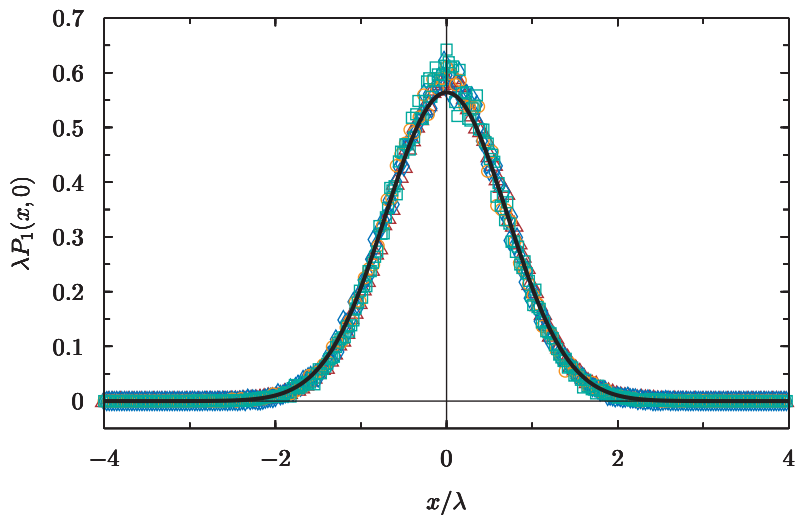}
\caption{Passage probability in rescaled units. Symbols correspond to the data sets: $T=2$, $R=201$, $L=352$ (brown $\triangle$), $T=2.1$, $R=321$, $L=602$ (blue $\diamond$), $T=2$, $R=161$, $L=302$ (orange $\varbigcirc$), $T=2.15$, $R=301$, $L=602$ (green $\square$). The black curve is the function $\pi^{-1/2} \textrm{e}^{-(x/\lambda)^{2}}$ given by (\ref{2026}).}
\label{fig_datacollapsepassageprobability}
\end{figure}

The passage probability $P_{1}(x,y=0)$ is thus extracted by sampling interfacial crossings along the line $y=0$ over a statistically adequate sample of MC snapshots. According to (\ref{2026}), data sets at different temperatures and system size collapse onto a Gaussian curve in a plot of $\lambda P_{1}(x,0)$ as function of $x/\lambda=\eta$; this is precisely what we observe in Fig.~\ref{fig_datacollapsepassageprobability}.
For the data sets in Fig.~\ref{fig_datacollapsepassageprobability} at the temperatures $T=2$, $T=2.1$, and $T=2.15$, we have sampled, respectively, $190513$, $54636$, and $62001$ single crossings on the $x$-axis. The Gaussian fluctuations exhibited by the interface are well reproduced by our data obtained for a sample of single crossings. It is important to mention that multiple crossings are inevitably observed on the lattice, while they do not appear in the probabilistic formulation. Multiple crossings are responsible for the occurrence of overhangs, as depicted in Fig.~ \ref{fig_real_medial_b}. It is actually possible either to restrict the sampling to those configurations with no multiple crossings, or to specify a certain rule for the treatment of multiple crossings. Different rules can be formulated; for instance, one can take the arithmetic average of the crossings abscissas. Once we have stipulated the sampling rule, we proceed with the construction of histograms which then we compare with the theoretical prediction (\ref{2026}). For the data sets of Fig.~\ref{fig_datacollapsepassageprobability} the statistics over configurations with either single or multiple crossings do not produce significant variations on the variance when comparing the theoretical prediction of (\ref{2026}).

\section{Two-point correlation functions}
\label{sec3}
The investigation carried out in the previous section is extended to pair correlation functions. In Sec.~\ref{sec3.1}, we compute the two-point correlation function of the energy density field within the exact field-theoretical approach. In Sec.~ \ref{sec3.2}, we show how the probabilistic interpretation can be extended in order to take into account energy density correlations. We then extract the passage probability which allows us to find exact expressions for the spin-spin correlation function. The results obtained within the probabilistic approach are identical to those obtained by means of the field-theoretic calculation of \cite{DS_twopoint}, as consistency requires.

\subsection{Energy density correlators}
\label{sec3.1}
We begin by computing the pair correlation function of the energy density field. The quantity we are interested in is defined by
\begin{equation}
\label{3001}
\langle \varepsilon(x_{1},y_{1}) \varepsilon(x_{2},y_{2}) \rangle_{-+} = \frac{ \langle B_{-+}(0;\textrm{i}R/2) \vert \varepsilon(x_{1},y_{1}) \varepsilon(x_{2},y_{2}) \vert B_{+-}(0;-\textrm{i}R/2) \rangle }{ \langle B_{-+}(0;\textrm{i}R/2) \vert B_{+-}(0;-\textrm{i}R/2) \rangle } \, ,
\end{equation}
with $-R/2<y_{2}<y_{1}<R/2$. In practice, we take $\xi \ll y_{1}-y_{2} \ll R$ and the distance from of the two fields from the boundaries to be large compared to $\xi$. Within these limits the expansion of the boundary state and the insertion of a complete set of multi-kink states between the two energy density fields is dominated by the single-kink state. The calculation proceeds as follows
\begin{equation}
\begin{aligned}
\label{3002}
\langle \varepsilon(x_{1},y_{1}) \varepsilon(x_{2},y_{2}) \rangle_{-+} & = \frac{1}{\mathcal{Z}_{-+}(R)}\int_{\mathbb{R}^{2}} \frac{\textrm{d}\theta_{1}\textrm{d}\theta_{2}\textrm{d}\theta_{3}}{(2\pi)^{3}} f_{-+}^{*}(\theta_{1}) f_{-+}(\theta_{3}) \mathcal{M}_{-+}^{\varepsilon}(\theta_{1}\vert\theta_{2}) \mathcal{M}_{-+}^{\varepsilon}(\theta_{2}\vert\theta_{3}) \times \\
& \times \mathcal{U}(\theta_{1},\theta_{2},\theta_{3}) \, ,
\end{aligned}
\end{equation}
where
\begin{align}\nonumber
\label{3003}
\mathcal{U}(\theta_{1},\theta_{2},\theta_{3}) & = \exp\Bigl[ - m(R/2-y_{1})\cosh\theta_{1} - m(y_{1}-y_{2})\cosh\theta_{2}-m(R/2+y_{2})\cosh\theta_{3} + \\
& + \textrm{i} mx(\sinh\theta_{1}-\sinh\theta_{2}) + \textrm{i} mx(\sinh\theta_{2}-\sinh\theta_{3}) \Bigr] \, .
\end{align}
For both matrix elements of the energy density field, we apply the decomposition (\ref{2014}) which entails a fully connected part (c), two partially connected ones and a fully disconnected one. The calculation gives
\begin{equation}
\label{3004}
\langle \varepsilon(x_{1},y_{1}) \varepsilon(x_{2},y_{2}) \rangle_{-+} = \langle \varepsilon(x_{1},y_{1}) \varepsilon(x_{2},y_{2}) \rangle_{-+}^{\textrm{c}} + \langle \varepsilon(x_{1},y_{1}) \rangle_{-+}^{\textrm{c}} \langle \varepsilon \rangle + \langle \varepsilon(x_{2},y_{2}) \rangle_{-+}^{\textrm{c}} \langle \varepsilon \rangle + \langle \varepsilon \rangle^{2} \, ,
\end{equation}
where $\langle \varepsilon(x_{j},y_{j}) \rangle_{-+}^{\textrm{c}} = \langle \varepsilon(x_{j},y_{j}) \rangle_{-+} - \langle \varepsilon \rangle$ is the connected part of the energy density profile. The connected component of the two-point correlation function is obtained from the product of form factors $F_{\varepsilon}(\theta_{12}+\textrm{i}\pi)F_{\varepsilon}(\theta_{23}+\textrm{i}\pi)$. A straightforward saddle-point calculation gives, at leading order,
\begin{equation}
\label{3005}
\langle \varepsilon(x_{1},y_{1}) \varepsilon(x_{2},y_{2}) \rangle_{-+}^{\textrm{c}} = \frac{F_{\varepsilon}^{2}(\textrm{i}\pi)}{m^{2}} P_{2}(x_{1},y_{1};x_{2},y_{2}) \, ,
\end{equation}
where $P_{2}$ is the function
\begin{equation}
\label{3006}
P_{2}(x_{1},y_{1};x_{2},y_{2}) = \frac{\textrm{e}^{-\frac{\eta_{1}^{2}}{2(1-\tau_{1})}-\frac{\eta_{2}^{2}}{2(1+\tau_{2})}-\frac{(\eta_{1}-\eta_{2})^{2}}{2(\tau_{1}-\tau_{2})}}}{\pi\lambda^{2}\sqrt{2(1-\tau_{1})(\tau_{1}-\tau_{2})(1+\tau_{2})}} \, .
\end{equation}
We note that $\langle \varepsilon(x_{j},y_{j}) \rangle_{-+}^{\textrm{c}}=O(R^{-1/2})$ at leading order and the next correction comes at order $O(R^{-3/2})$. The first term on the right hand side of (\ref{3004}) is proportional to $R^{-1}$ (see (\ref{3006})). It is then easy to check that the next correction in (\ref{3005}) comes at order $R^{-2}$. Summarizing, the energy density correlation function given in (\ref{3004}) is correct up to $O(R^{-3/2})$. Collecting the results obtained so far, the energy density correlation function including corrections at order $R^{-1}$ reads 
\begin{equation}
\label{3007}
\langle \varepsilon(x_{1},y_{1}) \varepsilon(x_{2},y_{2}) \rangle_{-+} = \frac{F_{\varepsilon}^{2}(\textrm{i}\pi)}{m^{2}} P_{2}(x_{1},y_{1};x_{2},y_{2}) + \langle \varepsilon \rangle \frac{F_{\varepsilon}(\textrm{i}\pi)}{m} \biggl[ P_{1}(x_{1},y_{1}) + P_{1}(x_{2},y_{2}) \biggr] + \langle \varepsilon \rangle^{2} \, .
\end{equation}
We observe that (\ref{3007}) satisfies the clustering properties
\begin{equation}
\begin{aligned}
\label{3008}
\lim_{x_{1} \rightarrow \pm \infty} \langle \varepsilon(x_{1},y_{1}) \varepsilon(x_{2},y_{2}) \rangle_{-+} & = \langle \varepsilon \rangle \langle \varepsilon(x_{2},y_{2}) \rangle_{-+} \, , \\
\lim_{x_{2} \rightarrow \pm \infty} \langle \varepsilon(x_{1},y_{1}) \varepsilon(x_{2},y_{2}) \rangle_{-+} & = \langle \varepsilon \rangle \langle \varepsilon(x_{1},y_{1}) \rangle_{-+} \, .
\end{aligned}
\end{equation}
In analogy with the energy density profile (\ref{2017}), energy density correlations show a non-trivial dependence on the coordinates only in the proximity of the interfacial region. This dependence is completely codified by (\ref{3007}).

The extension of the probabilistic interpretation to pair correlation functions reads
\begin{equation}
\label{3009}
\langle \varepsilon(x_{1},y_{1}) \varepsilon(x_{2},y_{2}) \rangle_{-+} = \int_{\mathbb{R}^{2}}\textrm{d}u_{1} \textrm{d}u_{2} \, \varepsilon(x_{1}\vert u_{1}) \varepsilon(x_{2}\vert u_{2}) P_{2}(u_{1},y_{1};u_{2},y_{2}) \, ,
\end{equation}
where $P_{2}(u_{1},y_{1};u_{2},y_{2})$ is the two-interval joint passage probability density. The quantity $P_{2}(u_{1},y_{1};u_{2},y_{2})\textrm{d}u_{1}\textrm{d}u_{2}$ is the net probability for the sharp interface to pass through the intervals $(u_{1},u_{1}+\textrm{d}u_{1})$ at ordinate $y_{1}$ \emph{and} $(u_{2},u_{2}+\textrm{d}u_{2})$ at ordinate $y_{2}$. It is indeed simple to show that (\ref{3009}) reproduces (\ref{3004}) with the passage probability density given by (\ref{3006}). Since $P_{2}$ is a joint probability distribution, by integrating over $x_{2}$ we obtain the marginal probability density
\begin{equation}
\label{3010}
P_{1}(x_{1},y_{1}) = \int_{\mathbb{R}}\textrm{d}x_{2} \, P_{2}(x_{1},y_{1};x_{2},y_{2})
\end{equation}
given by (\ref{2026}); an analogous relation follows by taking the marginal with respect to the other variable. A further integration leads to the normalization, i.e.
\begin{equation}
\label{3011}
\int_{\mathbb{R}^{2}}\textrm{d}x_{1}\textrm{d}x_{2} \, P_{2}(x_{1},y_{1};x_{2},y_{2}) = 1 \, .
\end{equation}

\subsection{Spin correlators}
\label{sec3.2}
We address the calculation of the spin-spin correlation function along the same lines outlined in Sec.~\ref{sec2}. The probabilistic reconstruction gives
\begin{equation}
\label{3012}
\langle \sigma(x_{1},y_{1}) \sigma(x_{2},y_{2}) \rangle_{-+} = \int_{\mathbb{R}^{2}}\textrm{d}u_{1} \textrm{d}u_{2} \, \sigma_{-+}(x_{1}\vert u_{1}) \sigma_{-+}(x_{2}\vert u_{2}) P_{2}(u_{1},y_{1};u_{2},y_{2}) \, ,
\end{equation}
where $\sigma(x_{j}\vert u_{j})$ is the sharp magnetization profile given by (\ref{2031}). Focusing on the leading-order term in the large-$R$ expansion, ignoring thus subleading terms coming from the interface structure, the twofold integral in (\ref{3012}) can be expressed in terms of cumulative distributions functions of the Gaussian distribution. To this end it is convenient to introduce the normal bivariate distribution
\begin{equation}
\label{3013}
\Pi_{2}(x_{1},x_{2}\vert\rho) = \frac{\textrm{e}^{-\frac{ x_{1}^{2} + x_{2}^{2} -2\rho x_{1}x_{2} }{2(1-\rho^{2})}}}{2\pi\sqrt{1-\rho^{2}}} \, ,
\end{equation}
where $\rho \in [0,1)$ is the correlation coefficient, i.e.
\begin{equation}
\label{3014}
\mathbb{E}[x_{i}x_{j}] = \int_{\mathbb{R}^{2}} \textrm{d}x_{1} \textrm{d}x_{2} \, x_{i} x_{j} \Pi_{2}(x_{1},x_{2}\vert\rho) = \delta_{ij} + (1-\delta_{ij})\rho \, ,  \qquad i,j \in \{1,2\} \, .
\end{equation}
The cumulative distribution function is therefore
\begin{equation}
\label{3015}
\Phi_{2}(x_{1},x_{2}\vert\rho) = \int_{-\infty}^{x_{1}}\textrm{d}u_{1} \int_{-\infty}^{x_{2}}\textrm{d}u_{2} \, \Pi_{2}(u_{1},u_{2}\vert\rho) \, .
\end{equation}
The passage probability can be expressed in terms of the normal bivariate distribution
\begin{equation}
\label{3016}
P_{2}(x_{1},y_{1};x_{2},y_{2}) = \frac{2}{\kappa_{1}\kappa_{2}\lambda^{2}} \Pi_{2}(\sqrt{2}\chi_{1},\sqrt{2}\chi_{2}\vert\rho)
\end{equation}
with $\chi_{j}=\eta_{j}/\kappa_{j}$, $\kappa_{j}=\sqrt{1-\tau_{j}^{2}}$, $\tau_{j}=2y_{j}/R$, $\eta_{j}=x_{j}/\lambda$, while the correlation coefficient reads
\begin{equation}
\label{3017}
\rho = \sqrt{\frac{1-\tau_{1}}{1+\tau_{1}} \frac{1+\tau_{2}}{1-\tau_{2}}} \, .
\end{equation}
It is worth noticing that the special limits $\rho \rightarrow 0$ (absence of correlations) and $\rho \rightarrow 1$ (perfect correlations) are never realized within the limits of validity of the field theoretical derivation because the two spin fields are both far from each other and far from the boundaries \cite{DS_twopoint}.

Thanks to the notions introduced above, the two-point correlation function (\ref{3012}) admits the following representation
\begin{equation}
\label{3018}
\langle \sigma(x_{1},y_{1}) \sigma(x_{2},y_{2}) \rangle_{-+}/M^{2} = 4 \Phi_{2}(\sqrt{2}\chi_{1},\sqrt{2}\chi_{2}\vert\rho) - 2\Phi_{1}(\sqrt{2}\chi_{1}) - 2\Phi_{1}(\sqrt{2}\chi_{2}) + 1 \, ,
\end{equation}
up to corrections due to interface structure which are computed in \cite{DS_twopoint}. Analogously, the one-point correlation function of the spin field can be written as follows
\begin{equation}
\label{3019}
\langle \sigma(x,y) \rangle_{-+}/M = 2 \Phi_{1}(\sqrt{2}\chi) - 1 \, ,
\end{equation}
where $\Phi_{1}(x)$ is the cumulative distribution of the standardized Gaussian probability distribution, i.e.
\begin{equation}
\label{3020}
\Phi_{1}(x) = \int_{-\infty}^{x}\textrm{d}u \, \Pi_{1}(u) = \frac{1+\textrm{erf}(x/\sqrt{2})}{2}\, , 
\end{equation}
with
\begin{equation}
\label{3020_a}
\Pi_{1}(x) = \frac{\textrm{e}^{-x^{2}/2}}{\sqrt{2\pi}} \, .
\end{equation}
In fact, (\ref{3019}) with $y=0$ is nothing but the leading order term in (\ref{2034}).

By using the following properties of cumulative distribution functions
\begin{equation}
\begin{aligned}
\label{3021}
\lim_{x_{1}\rightarrow -\infty} \Phi_{2}(x_{1},x_{2} \vert \rho) & = 0 \, , \\
\lim_{x_{2}\rightarrow -\infty} \Phi_{2}(x_{1},x_{2} \vert \rho) & = 0 \, , \\
\lim_{x_{1}\rightarrow +\infty} \Phi_{1}(x_{1},x_{2} \vert \rho) & = \Phi_{1}(x_{2}) \, , \\
\lim_{x_{2}\rightarrow +\infty} \Phi_{2}(x_{1},x_{2} \vert \rho) & = \Phi_{1}(x_{1}) \, ,
\end{aligned}
\end{equation}
we can derive the following clustering properties of the spin-spin correlation function
\begin{equation}
\begin{aligned}
\label{3022}
\lim_{x_{1} \rightarrow \pm \infty} \langle \sigma(x_{1},y_{1}) \sigma(x_{2},y_{2}) \rangle_{-+} & = \pm M \langle \sigma(x_{2},y_{2}) \rangle_{-+} \, , \\
\lim_{x_{2} \rightarrow \pm \infty} \langle \sigma(x_{1},y_{1}) \sigma(x_{2},y_{2}) \rangle_{-+} & = \pm M \langle \sigma(x_{1},y_{1}) \rangle_{-+} \, ,
\end{aligned}
\end{equation}
which are actually a particular case of Eq. (3.18) in \cite{DS_twopoint}.

\subsection{Numerical results}
\begin{figure*}[htbp]
\centering
        \begin{subfigure}[b]{0.3\textwidth}
            \centering
\begin{tikzpicture}[thick, line cap=round, >=latex, scale=0.22]
\tikzset{fontscale/.style = {font=\relsize{#1}}}
\draw[thin, dashed, -] (-10, 0) -- (11, 0) node[below] {};
\draw[thin, dashed, -] (0, -5) -- (0, 5) node[left] {};
\draw[very thick, red, -] (0, 5) -- (10, 5) node[] {};
\draw[very thick, red, -] (0, -5) -- (10, -5) node[] {};
\draw[very thick, blue, -] (-10, 5) -- (0, 5) node[] {};
\draw[very thick, blue, -] (-10, -5) -- (0, -5) node[] {};
\draw[thin, fill=green!30] (4, 3.3) circle (10pt) node[right] { $(x,y)$ };;
\draw[thin, fill=green!30] (4, -3.3) circle (10pt) node[right] { $(x,-y)$ };;
\draw[thin, fill=white] (-5, 5) circle (0pt) node[above] {${\color{blue}{-}}$};;
\draw[thin, fill=white] (-5, -5) circle (0pt) node[below] {${\color{blue}{-}}$};;
\draw[thin, fill=white] (5, 5) circle (0pt) node[above] {${\color{red}{+}}$};;
\draw[thin, fill=white] (5, -5) circle (0pt) node[below] {${\color{red}{+}}$};;
\end{tikzpicture}
            \caption[]%
            {{\small $\mathcal{G}_{\textrm{v}}^{\sigma\sigma}(x,y)$}}    
            \label{fig_ssv}
        \end{subfigure}
\hfill
        \begin{subfigure}[b]{0.3\textwidth}
            \centering
\begin{tikzpicture}[thick, line cap=round, >=latex, scale=0.22]
\tikzset{fontscale/.style = {font=\relsize{#1}}}
\draw[thin, dashed, -] (-10, 0) -- (11, 0) node[below] {};
\draw[thin, dashed, -] (0, -5) -- (0, 5) node[left] {};
\draw[very thick, red, -] (0, 5) -- (10, 5) node[] {};
\draw[very thick, red, -] (0, -5) -- (10, -5) node[] {};
\draw[very thick, blue, -] (-10, 5) -- (0, 5) node[] {};
\draw[very thick, blue, -] (-10, -5) -- (0, -5) node[] {};
\draw[thin, fill=green!30] (2, 3.3) circle (10pt) node[right] { $(x,y)$ };;
\draw[thin, fill=green!30] (-2, -3.3) circle (10pt) node[left] { $(-x,-y)$ };;
\draw[thin, fill=white] (-5, 5) circle (0pt) node[above] {${\color{blue}{-}}$};;
\draw[thin, fill=white] (-5, -5) circle (0pt) node[below] {${\color{blue}{-}}$};;
\draw[thin, fill=white] (5, 5) circle (0pt) node[above] {${\color{red}{+}}$};;
\draw[thin, fill=white] (5, -5) circle (0pt) node[below] {${\color{red}{+}}$};;
\end{tikzpicture}
            \caption[]%
            {{\small $\mathcal{G}_{\textrm{t}}^{\sigma\sigma}(x,y)$}}    
            \label{fig_sst}
        \end{subfigure}
\hfill
        \begin{subfigure}[b]{0.3\textwidth}
            \centering
\begin{tikzpicture}[thick, line cap=round, >=latex, scale=0.22]
\tikzset{fontscale/.style = {font=\relsize{#1}}}
\draw[thin, dashed, -] (-10, 0) -- (11, 0) node[below] {};
\draw[thin, dashed, -] (0, -5) -- (0, 5) node[left] {};
\draw[very thick, red, -] (0, 5) -- (10, 5) node[] {};
\draw[very thick, red, -] (0, -5) -- (10, -5) node[] {};
\draw[very thick, blue, -] (-10, 5) -- (0, 5) node[] {};
\draw[very thick, blue, -] (-10, -5) -- (0, -5) node[] {};
\draw[thin, fill=green!30] (0, 3.3) circle (10pt) node[right] { $(0,y_{1})$ };;
\draw[thin, fill=green!30] (0, -1.5) circle (10pt) node[right] { $(0,y_{2})$ };;
\draw[thin, fill=white] (-5, 5) circle (0pt) node[above] {${\color{blue}{-}}$};;
\draw[thin, fill=white] (-5, -5) circle (0pt) node[below] {${\color{blue}{-}}$};;
\draw[thin, fill=white] (5, 5) circle (0pt) node[above] {${\color{red}{+}}$};;
\draw[thin, fill=white] (5, -5) circle (0pt) node[below] {${\color{red}{+}}$};;
\end{tikzpicture}
            \caption[]%
            {{\small $\mathcal{G}_{\textrm{i}}^{\sigma\sigma}(y_{1},y_{2})$}}    
            \label{fig_ssi}
        \end{subfigure}
\vskip\baselineskip
\caption[]
{\small The spin-spin correlation functions considered in the text: vertical alignment (a), tilted alignment (b), alignment along the interface (c).}
\label{fig_ss}
\end{figure*}
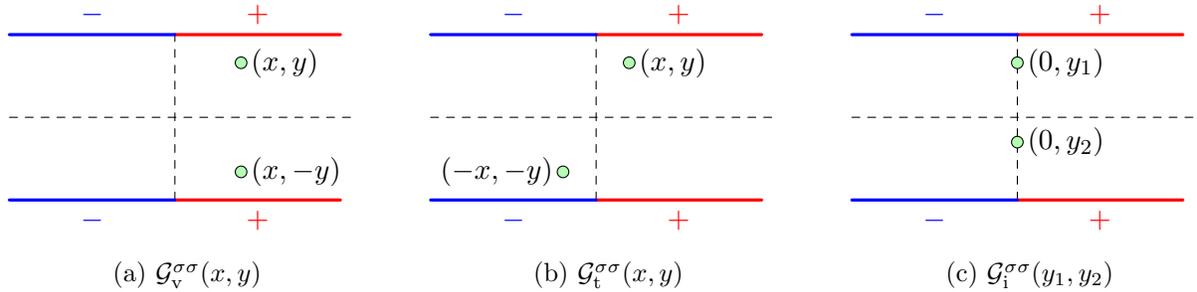

In the following, we specialize the general result (\ref{3018}) to a certain class of configurations in which the spin fields are arranged in a symmetric fashion. The configurations we are going to examine are those depicted in Fig.~ \ref{fig_ss}.

As illustrated in \cite{DS_twopoint}, the formal result provided by (\ref{3018}) admits a more explicit formulation in terms of Owen's $T$-function \cite{Owen1956,Owen1980}, whose definition and main properties are collected in Appendix \ref{appendixB}. The analytic expressions of the spin-spin correlation functions for the correlators illustrated in Fig.~\ref{fig_ss} are given by:
\begin{equation}
\begin{aligned}
\label{3023}
\mathcal{G}_{\textrm{v}}^{\sigma\sigma}(x,y) & = \langle \sigma(x,y) \sigma(x,-y) \rangle_{-+}/M^{2} = 1 - 8T(\sqrt{2}\chi,\sqrt{\tau}) \, , \\
\mathcal{G}_{\textrm{t}}^{\sigma\sigma}(x,y) & = \langle \sigma(x,y) \sigma(-x,-y) \rangle_{-+}/M^{2} = 8T(\sqrt{2}\chi,1/\sqrt{\tau}) - 1 \, , \\
\mathcal{G}_{\textrm{i}}^{\sigma\sigma}(y_{1},y_{2}) & = \langle \sigma(0,y_{1}) \sigma(0,y_{2}) \rangle_{-+}/M^{2} = \frac{2}{\pi} \tan^{-1} \sqrt{\frac{(1-\tau_{1})(1+\tau_{2})}{2(\tau_{1}-\tau_{2})}}\, ,
\end{aligned}
\end{equation}
corresponding respectively to the vertical alignment ($\textrm{v}$), tilted alignment ($\textrm{t}$), and alignment along the interface support ($\textrm{i}$). In the above, $\chi=\eta/\sqrt{1-\tau^{2}}$, $\eta=x/\lambda$, and $\tau=2y/R$.

We now provide the comparison between the analytical results (\ref{3023}) and the numerical simulations we performed. The correlation function $\mathcal{G}_{\textrm{v}}^{\sigma\sigma}(x,y)$ is plotted in Fig.~\ref{fig_ssv} as a function of $x$ for several values of $y$. We have restricted the plot to positive values of $x$ because $\mathcal{G}_{\textrm{v}}^{\sigma\sigma}(x,y)$ is an even function of $x$.
\begin{figure}[htbp]
\centering
\includegraphics[width=105mm]{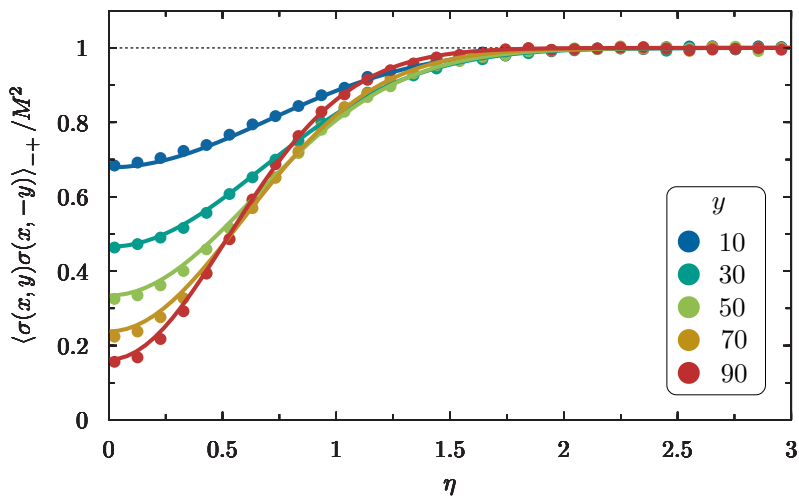}
\caption{Vertical correlation function $\mathcal{G}_{\textrm{v}}^{\sigma\sigma}(x,y)$ as a function of $x$ for the values of $y$ indicated in the inset. Numerical data are obtained from MC simulations with $T=2.15$, $L=602$, $R=301$. Solid curves correspond to the analytic result given in (\ref{3023}).}
\label{fig_ssv}
\end{figure}
Far away from the interfacial region, i.e., $|\eta|=|x|/\lambda \gg 1$, the correlation function approaches the square of the spontaneous magnetization, as required by the clustering properties; thus, $\mathcal{G}_{\textrm{v}}^{\sigma\sigma}(x \rightarrow +\infty,y) \rightarrow +1$.

Within the probabilistic interpretation, the configurations in which the interface reaches the two spins occur rarely when $|x| \gg \lambda$. Then, from the sharp profile (\ref{2031}) it follows that each spin field ``carries'' a factor $M$, thus $\mathcal{G}_{\textrm{v}}^{\sigma\sigma}$ approaches $1$. In the closeness of the interfacial region instead the correlation is less than $M^{2}$; thus, $0<\mathcal{G}_{\textrm{v}}^{\sigma\sigma}<1$. The occurrence of such a feature is easily interpreted within the probabilistic picture. Configurations in which the sharp interface passes between the two spins are weighted with the negative factor equal to $-M^{2}$; it thus follows how the correlation decreases with respect to the far right/left regions. Analogously, the increase of the correlation function upon decreasing $y$ at $x=0$ can be interpreted by reasoning along the same lines. For small $y$ the two spin fields come closer and configurations in which the interface passes through them are less probable, consequently, the correlation increases. All the features described above are reproduced by numerical data and are captured by the analytic result, as illustrated in Fig.~\ref{fig_ssv}.

The numerical results for the tilted correlation function $\mathcal{G}_{\textrm{t}}^{\sigma\sigma}(x,y)$ are provided in Fig.~\ref{fig_sst} together with the analytic results. Even in this case the correlation function is an even function of $x$.
\begin{figure}[htbp]
\centering
\includegraphics[width=105mm]{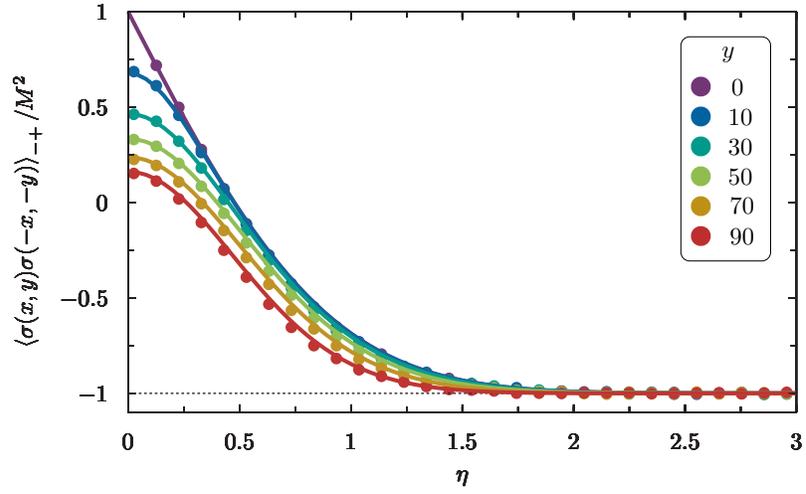}
\caption{Tilted correlation function $\mathcal{G}_{\textrm{t}}^{\sigma\sigma}(x,y)$ as a function of $x$ for the values of $y$ indicated in the inset. Numerical data are obtained from MC simulations with $T=2.15$, $L=602$, $R=301$. Solid curves correspond to the analytic result given in (\ref{3023}).}
\label{fig_sst}
\end{figure}
The asymptotic behavior is again straightforwardly interpreted within the probabilistic picture. Far away from the interface the two spins field probe two opposite phases, therefore the correlation function approaches $-M^{2}$, correspondingly $\mathcal{G}_{\textrm{t}}^{\sigma\sigma}\rightarrow -1$. In the limit $x\rightarrow0$ the tilted and the vertical configurations degenerate onto the same one; hence, the interpretation followed for the vertical configuration applies also to the tilted one. Surprisingly enough, the numerical data follow rather accurately the analytical result also in the limit $\tau \rightarrow0$ ($\rho \rightarrow 1$) which is not covered by the domain of validity of the theory for small $x$. Such a limit corresponds to horizontally aligned spins in positions $(-x,0)$ and $(0,x)$. For this special limit the correlation simplifies as follows $\mathcal{G}_{\textrm{t}}^{\sigma\sigma}(x,y\rightarrow0) \rightarrow 1-2\textrm{erf}(|\eta|)$; see the solid purple line in Fig.~\ref{fig_sst}.

In Fig.~\ref{fig_ssi}, we compare the numerical and analytical results for the correlation function along the interface with spins fields equally spaced with respect to the $x$ axis, i.e., $\mathcal{G}_{\textrm{i}}^{\sigma\sigma}(y,-y)$.
\begin{figure}[htbp]
\centering
\includegraphics[width=105mm]{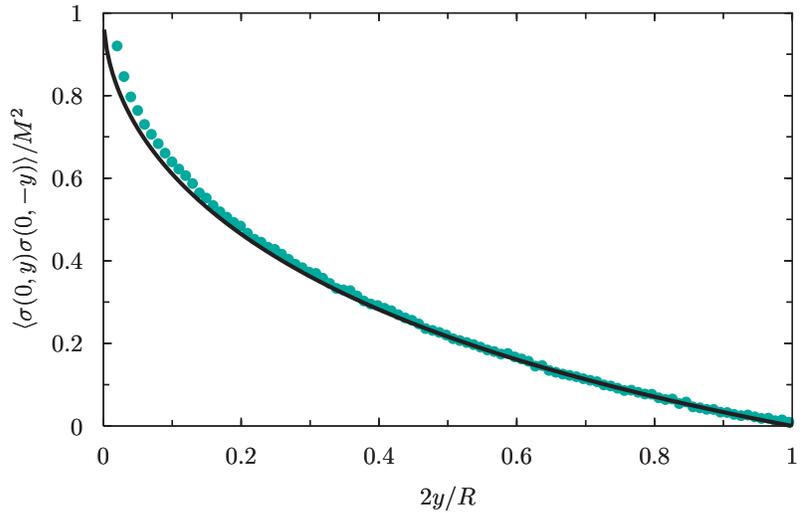}
\caption{Correlation function along the interface $\mathcal{G}_{\textrm{i}}^{\sigma\sigma}(y,-y)$. Numerical data (green dots) are obtained from MC simulations with $T=2.15$, $L=602$, $R=301$. The solid black line corresponds to the analytic result (\ref{3024}).}
\label{fig_ssi}
\end{figure}
For such a specific configuration, $y=y_{1}=-y_{2}$, and the analytic result gives\footnote{We recall the identity (\ref{Owen_special_1}).}
\begin{equation}
\begin{aligned}
\label{3024}
\mathcal{G}_{\textrm{i}}^{\sigma\sigma}(y,-y) & = \frac{2}{\pi} \tan^{-1}\frac{1-\tau}{2\sqrt{\tau}} \, , \\
& = 1 - \frac{4}{\pi} \tan^{-1}\sqrt{\tau} \, , \\
& = \frac{2}{\pi}\sin^{-1}\frac{1-\tau}{1+\tau} \, .  
\end{aligned}
\end{equation}
The long-range form of interfacial correlations can be visualized in a direct fashion by expanding (\ref{3024}) for small $\tau=2y/R$; hence,
\begin{equation}
\begin{aligned}
\label{}
\mathcal{G}_{\textrm{i}}^{\sigma\sigma}(y,-y) & = 1 - 4\sqrt{\frac{2y}{\pi R}} + O((y/R)^{3/2}) \, .
\end{aligned}
\end{equation}
The power law behavior proportional to $\sqrt{y/R}$ is the signature of long range correlations mediated by the interface. A good agreement between theory and numerics is observed for a wide range of $y \in (0,R/2)$ with deviations for $y \rightarrow 0$. The deviations from the analytic result, which is obtained by exploiting the infrared properties of the single-particle contribution, are expected to occur when the assumption $y_{1}-y_{2} \gg \xi$ is violated.

\section{Three-point correlation functions}
\label{sec4}
In the previous sections, we have showed how one- and two-point correlation functions of both the spin and energy density fields can be obtained within a probabilistic formulation in which the passage probability follows from the field-theoretical calculation of energy density correlations. The results obtained for one- and two-point correlation functions of the spin field agree with those obtained directly from field theory, respectively in \cite{DV} for magnetization profiles and in \cite{DS_twopoint} for spin-spin correlation functions. The logic discussed above applies to the three-point correlation functions discussed in this section, and more generally to arbitrary $n$-point correlation functions \cite{Squarcini_Multipoint}. In particular, the fully connected part of energy density correlations is proportional to the passage probability, a feature that we have shown explicitly for $n=1$ and $n=2$. By following the strategy summarized above, we compute three-point correlation functions of the spin field in various arrangements. The occurrence of long range interfacial correlations and their explicit form is also examined.

\subsection{Energy density correlators}
By following the guidelines outlined in Secs.~\ref{sec2} and \ref{sec3}, we commence by computing the three-point energy density correlation function. The object of our interest is thus
\begin{equation}
\label{4001}
\langle \varepsilon(x_{1},y_{1}) \varepsilon(x_{2},y_{2}) \varepsilon(x_{3},y_{3}) \rangle_{-+} = \frac{ \langle B_{-+}(0;\textrm{i}R/2) \vert \varepsilon(x_{1},y_{1}) \varepsilon(x_{2},y_{2}) \varepsilon(x_{3},y_{3}) \vert B_{+-}(0;-\textrm{i}R/2) \rangle }{ \langle B_{-+}(0;\textrm{i}R/2) \vert B_{+-}(0;-\textrm{i}R/2) \rangle } \, ,
\end{equation}
with the ordering $y_{1}>y_{2}>y_{3}$ and large separation between fields and boundaries is also assumed. Since the calculation of (\ref{4001}) follows precisely the same path detailed in Sec.~\ref{sec3}, we can skip some intermediate steps and present the final result for the connected part of (\ref{4001}), which reads
\begin{equation}
\label{4002}
\langle \varepsilon(x_{1},y_{1}) \varepsilon(x_{2},y_{2}) \varepsilon(x_{3},y_{3}) \rangle_{-+}^{\textrm{c}} = \frac{ F_{\varepsilon}^{3}(\textrm{i}\pi) }{ m^{3} } P_{3}(x_{1},y_{1};x_{2},y_{2};x_{3},y_{3}) \, ,
\end{equation}
with $P_{3}$ the three-intervals joint passage probability density. The contributions stemming from the disconnected pieces can be taken into account by extending the arguments of Sec.~\ref{sec3}. Analogously to the two-interval case, $P_{3}$ can be expressed in terms of the trivariate normal distribution $\Pi_{3}(u_{1},u_{2},u_{3} \vert \rho_{12}, \rho_{13}, \rho_{23})$, therefore
\begin{equation}
\label{4003}
P_{3}(x_{1},y_{1};x_{2},y_{2};x_{3},y_{3}) = \frac{2^{3/2}}{\kappa_{1}\kappa_{2}\kappa_{3}\lambda^{3}} \Pi_{3}(\sqrt{2}\chi_{1},\sqrt{2}\chi_{2},\sqrt{2}\chi_{3} \vert \rho_{12}, \rho_{13}, \rho_{23}) \, ,
\end{equation}
with correlation coefficients
\begin{equation}
\label{4004}
\rho_{ij} = \sqrt{\frac{ 1-\tau_{i} }{ 1+\tau_{i} } \frac{ 1+\tau_{j} }{ 1-\tau_{j} }} \, 
\end{equation}
for $i<j$. Notice that only two of the above coefficients are independent by virtue of the Markov property $\rho_{13}=\rho_{12}\rho_{23}$.

\subsection{Spin field correlators}
\label{sec42}
Once we have determined the passage probability, we can apply the probabilistic framework in order to compute the three-point correlation function of the spin field. We have
\begin{equation}
\label{4005}
\langle \sigma(x_{1},y_{1}) \sigma(x_{2},y_{2}) \sigma(x_{3},y_{3}) \rangle_{-+} = \int_{\mathbb{R}^{3}}\textrm{d}u_{1}\textrm{d}u_{2}\textrm{d}u_{3} \, P_{3}(u_{1},y_{1};u_{2},y_{2};u_{3},y_{3}) \prod_{j=1}^{3} \sigma_{-+}(x_{j}\vert u_{j}) \, ,
\end{equation}
with sharp profiles given by (\ref{2031}). The calculation of (\ref{4005}) follows from a simple extension of (\ref{3018}) which, for the case at hand, it reads
\begin{equation}
\begin{aligned}
\label{4006}
\langle \sigma(x_{1},y_{1}) \sigma(x_{2},y_{2}) \sigma(x_{3},y_{3}) \rangle_{-+}/M^{3} & = 8 \Phi_{3}(\sqrt{2}\chi_{1},\sqrt{2}\chi_{2},\sqrt{2}\chi_{3} \vert \rho_{12},\rho_{13},\rho_{23}) + \\
& - 4 \Phi_{2}(\sqrt{2}\chi_{1},\sqrt{2}\chi_{2} \vert \rho_{12}) - 4 \Phi_{2}(\sqrt{2}\chi_{1},\sqrt{2}\chi_{3} \vert \rho_{13}) + \\
& - 4 \Phi_{2}(\sqrt{2}\chi_{2},\sqrt{2}\chi_{3} \vert \rho_{23}) + 2 \Phi_{1}(\sqrt{2}\chi_{1}) + 2 \Phi_{1}(\sqrt{2}\chi_{2}) + \\
& + 2 \Phi_{1}(\sqrt{2}\chi_{3}) -1 \, ,
\end{aligned}
\end{equation}
where
\begin{equation}
\label{ }
\Phi_{3}(x_{1}, x_{2}, x_{3} \vert \rho_{12},\rho_{13},\rho_{23}) = \int_{-\infty}^{x_{1}}\textrm{d}u_{1} \int_{-\infty}^{x_{2}}\textrm{d}u_{2} \int_{-\infty}^{x_{3}}\textrm{d}u_{3} \, \Pi_{3}(u_{1},u_{2},u_{3} \vert \rho_{12}, \rho_{13}, \rho_{23})
\end{equation}
is the cumulative distribution of the trivariate normal distribution $\Pi_{3}$, whose explicit expression is supplied in (\ref{B01_00}).

It is instructive to comment on some general properties of (\ref{4006}) before passing to a detailed examination of specific results. Firstly, we observe the clustering property
\begin{equation}
\label{4007}
\lim_{x_{3} \rightarrow \pm \infty} \langle \sigma(x_{1},y_{1}) \sigma(x_{2},y_{2}) \sigma(x_{3},y_{3}) \rangle_{-+} = \pm M \langle \sigma(x_{1},y_{1}) \sigma(x_{2},y_{2}) \rangle_{-+} \, ,
\end{equation}
and the analogous relations in which either $x_{1}$ or $x_{2}$ are sent to infinity. The relation (\ref{4007}) is a direct consequence of the asymptotic properties satisfied by cumulative distribution functions. Analogously, upon sending the three spin fields towards $\pm \infty$ with their relative separation kept fixed, (\ref{4005}) gives $\pm M^{3}$. Of course, the above fact follows because (\ref{4005}) contains the interfacial contribution of the three-point correlation function. The bulk contributions originate subleading corrections because they involve a higher number of intermediate states \cite{DS_twopoint}.

It is also interesting to observe how the three-point correlation function (\ref{4006}) vanishes when the three spin fields are placed along the straight line which joins the pinning points; along such a line, $x_{1}=x_{2}=x_{3}=0$. In order to prove this result it is enough to recall the quadrant probability, i.e., the probability of having $x_{1}<0$ and $x_{2}<0$ for the bivariate normal distribution (\ref{3013})
\begin{equation}
\label{4008}
\Phi_{2}(0,0 \vert \rho_{ij}) = \frac{1}{4}+\frac{1}{2\pi}\sin^{-1}\rho_{ij} \, ,
\end{equation}
and the orthant probability for the trivariate normal distribution
\begin{equation}
\label{4009}
\Phi_{3}(0,0,0 \vert \rho_{12},\rho_{13},\rho_{23}) = \frac{1}{8}+\frac{1}{4\pi}\left( \sin^{-1}\rho_{12}+\sin^{-1}\rho_{13}+\sin^{-1}\rho_{23} \right) \, .
\end{equation}
By plugging (\ref{4008}) and (\ref{4009}) into (\ref{4006}), we find
\begin{equation}
\label{0310}
\langle \sigma(0,y_{1}) \sigma(0,y_{2}) \sigma(0,y_{3}) \rangle_{-+} = 0 \, ,
\end{equation}
thus the correlation function vanishes irrespectively of $y_{1}$, $y_{2}$, and $y_{3}$. It is also possible to show how the three-point correlation function with spin fields arranged with a central symmetry has to vanish, i.e., $\langle \sigma(x,y) \sigma(0,0) \sigma(-x,-y) \rangle_{-+}=0$ for any $x$ and $y$.

\subsection{Symmetric configurations}
\label{secthreespincorrelators}
\begin{figure*}[htbp]
\centering
        \begin{subfigure}[b]{0.3\textwidth}
            \centering
\begin{tikzpicture}[thick, line cap=round, >=latex, scale=0.24]
\tikzset{fontscale/.style = {font=\relsize{#1}}}
\draw[thin, dashed, -] (-10, 0) -- (11, 0) node[below] {};
\draw[thin, dashed, -] (0, -5) -- (0, 5) node[left] {};
\draw[very thick, red, -] (0, 5) -- (10, 5) node[] {};
\draw[very thick, red, -] (0, -5) -- (10, -5) node[] {};
\draw[very thick, blue, -] (-10, 5) -- (0, 5) node[] {};
\draw[very thick, blue, -] (-10, -5) -- (0, -5) node[] {};
\draw[thin, fill=green!30] (0, 3.3) circle (10pt) node[left] { $(0,y)$ };;
\draw[thin, fill=green!30] (5, 0) circle (10pt) node[above] { $(x,0)$ };;
\draw[thin, fill=green!30] (0, -3.3) circle (10pt) node[left] { $(0,-y)$ };;
\draw[thin, fill=white] (-5, 5) circle (0pt) node[above] {${\color{blue}{-}}$};;
\draw[thin, fill=white] (-5, -5) circle (0pt) node[below] {${\color{blue}{-}}$};;
\draw[thin, fill=white] (5, 5) circle (0pt) node[above] {${\color{red}{+}}$};;
\draw[thin, fill=white] (5, -5) circle (0pt) node[below] {${\color{red}{+}}$};;
\end{tikzpicture}
            \caption[]%
            {{\small $\mathcal{G}_{A}^{\sigma\sigma\sigma}(x,y)$}}    
            \label{fig_sssv}
        \end{subfigure}
\hfill
        \begin{subfigure}[b]{0.3\textwidth}
            \centering
\begin{tikzpicture}[thick, line cap=round, >=latex, scale=0.24]
\tikzset{fontscale/.style = {font=\relsize{#1}}}
\draw[thin, dashed, -] (-10, 0) -- (-4, 0) node[below] {};
\draw[thin, dashed, -] (-0.5, 0) -- (10, 0) node[below] {};
\draw[thin, dashed, -] (0, -5) -- (0, 5) node[left] {};
\draw[very thick, red, -] (0, 5) -- (10, 5) node[] {};
\draw[very thick, red, -] (0, -5) -- (10, -5) node[] {};
\draw[very thick, blue, -] (-10, 5) -- (0, 5) node[] {};
\draw[very thick, blue, -] (-10, -5) -- (0, -5) node[] {};
\draw[thin, fill=green!30] (7, 3.3) circle (10pt) node[left] { $(x,y)$ };;
\draw[thin, fill=green!30] (0, 0) circle (10pt) node[left] { $(0,0)$ };;
\draw[thin, fill=green!30] (7, -3.3) circle (10pt) node[left] { $(x,-y)$ };;
\draw[thin, fill=white] (-5, 5) circle (0pt) node[above] {${\color{blue}{-}}$};;
\draw[thin, fill=white] (-5, -5) circle (0pt) node[below] {${\color{blue}{-}}$};;
\draw[thin, fill=white] (5, 5) circle (0pt) node[above] {${\color{red}{+}}$};;
\draw[thin, fill=white] (5, -5) circle (0pt) node[below] {${\color{red}{+}}$};;
\end{tikzpicture}
            \caption[]%
            {{\small $\mathcal{G}_{B}^{\sigma\sigma\sigma}(x,y)$}}    
            \label{fig_ssst}
        \end{subfigure}
\hfill
        \begin{subfigure}[b]{0.3\textwidth}
            \centering
\begin{tikzpicture}[thick, line cap=round, >=latex, scale=0.24]
\tikzset{fontscale/.style = {font=\relsize{#1}}}
\draw[thin, dashed, -] (-10, 0) -- (2.5, 0) node[below] {};
\draw[thin, dashed, -] (6.5, 0) -- (10, 0) node[below] {};
\draw[thin, dashed, -] (0, -5) -- (0, 5) node[left] {};
\draw[very thick, red, -] (0, 5) -- (10, 5) node[] {};
\draw[very thick, red, -] (0, -5) -- (10, -5) node[] {};
\draw[very thick, blue, -] (-10, 5) -- (0, 5) node[] {};
\draw[very thick, blue, -] (-10, -5) -- (0, -5) node[] {};
\draw[thin, fill=green!30] (7, 3.3) circle (10pt) node[left] { $(x,y)$ };;
\draw[thin, fill=green!30] (7, 0) circle (10pt) node[left] { $(x,0)$ };;
\draw[thin, fill=green!30] (7, -3.3) circle (10pt) node[left] { $(x,-y)$ };;
\draw[thin, fill=white] (-5, 5) circle (0pt) node[above] {${\color{blue}{-}}$};;
\draw[thin, fill=white] (-5, -5) circle (0pt) node[below] {${\color{blue}{-}}$};;
\draw[thin, fill=white] (5, 5) circle (0pt) node[above] {${\color{red}{+}}$};;
\draw[thin, fill=white] (5, -5) circle (0pt) node[below] {${\color{red}{+}}$};;
\end{tikzpicture}
            \caption[]%
            {{\small $\mathcal{G}_{C}^{\sigma\sigma\sigma}(x,y)$}}    
            \label{fig_sssi}
        \end{subfigure}
\vskip\baselineskip
\caption[]
{\small The three point correlation functions of the spin field considered in the paper.t}
\label{fig_sss}
\end{figure*}
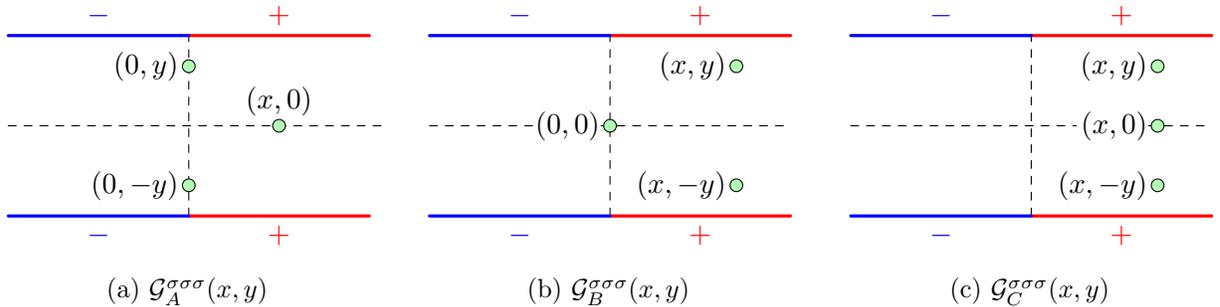

We specialize the general result (\ref{4006}) to the symmetric configurations in which the three spin fields are arranged as depicted in Fig.~\ref{fig_sss}.
The three-point correlation functions summarized in Fig.~\ref{fig_sss} are defined by:
\begin{equation}
\begin{aligned}
\label{11022021_1325}
\mathcal{G}_{A}^{\sigma\sigma\sigma}(x,y) & = \langle \sigma(0,y) \sigma(x,0) \sigma(0,-y) \rangle_{-+}/M^{3}  \\
\mathcal{G}_{B}^{\sigma\sigma\sigma}(x,y) & = \langle \sigma(x,y) \sigma(0,0) \sigma(x,-y) \rangle_{-+}/M^{3}  \\
\mathcal{G}_{C}^{\sigma\sigma\sigma}(x,y) & = \langle \sigma(x,y) \sigma(x,0) \sigma(x,-y) \rangle_{-+}/M^{3} \, .
\end{aligned}
\end{equation}
The manipulations which allowed us to express the two-point correlation function (\ref{3018}) into the form (\ref{3023}) can be applied -- mutatis mutandis -- to the three-point correlation functions (\ref{11022021_1325}). It is indeed possible to express the cumulative distribution $\Phi_{3}$ of the trivariate normal Gaussian in terms of Owen's $T$ \cite{Owen1980} and Steck's $S$ functions \cite{Steck1958}. Leaving in Appendix \ref{appendixB} the technicalities involved in such manipulations, here we simply quote the final results for the correlators of Fig.~\ref{fig_sss} and their comparison with MC simulations.\\

\subsubsection*{Configuration A}
For the configurations showed in Fig.~\ref{fig_sss} the correlation coefficients are given by $\rho_{12}=\rho_{23}=\sqrt{\rho_{13}}=\sqrt{(1-\tau)/(1+\tau)}\equiv\varrho$, with $\tau=2y/R$. The analytic expression for the correlation function reads
\begin{equation}
\label{C213}
\mathcal{G}_{A}^{\sigma\sigma\sigma}(x,y) = \frac{2}{\sqrt{\pi}} \int_{0}^{\eta}\textrm{d}u \, \textrm{erf}^{\, 2}(ru) \textrm{e}^{-u^{2}} \, ,
\end{equation}
with
\begin{equation}
\label{ }
r = \frac{\varrho}{\sqrt{1-\varrho^{2}}} = \sqrt{\frac{1-\tau}{2\tau}} \, .
\end{equation}
The symmetry property $\mathcal{G}_{A}^{\sigma\sigma\sigma}(x,y)=-\mathcal{G}_{A}^{\sigma\sigma\sigma}(-x,y)$ is manifest. This property is required by the anti-symmetry under parity, i.e., reversing the sign of $x$ corresponds to swap the $+$ and $-$ boundary conditions.

Let us discuss some general properties of (\ref{C213}). For fixed $y$ the correlator $\mathcal{G}_{A}^{\sigma\sigma\sigma}(x,y)$ is a monotonically increasing function of $x$. The asymptotic value attained for $x\rightarrow\pm\infty$ follows from the clustering property
\begin{equation}
\label{C211}
\lim_{x \rightarrow \pm \infty} \mathcal{G}_{A}^{\sigma\sigma\sigma}(x,y) = \pm \mathcal{G}_{\textrm{i}}^{\sigma\sigma}(y,-y) \, ,
\end{equation}
with the right hand side given by the two-point correlation function along the interface; see (\ref{3023}). In order to check (\ref{C211}) it is useful to use the following identities
\begin{equation}
\begin{aligned}
\label{13042021_2331}
\frac{2}{\sqrt{\pi}} \int_{0}^{\infty}\textrm{d}u \, \textrm{erf}^{\, 2}(ru) \textrm{e}^{-u^{2}} & = -1 + \frac{4}{\pi} \tan^{-1}\sqrt{1+2r^{2}} \\
& = \frac{2}{\pi}\sin^{-1}\frac{1-\tau}{1+\tau} \, ,
\end{aligned}
\end{equation}
which, thanks to (\ref{3024}), establishes the clustering identity (\ref{C211}).

For arbitrary values of $\tau$ the integral in (\ref{C213}) cannot be expressed in terms of elementary functions. However, for $\tau=1/3$, corresponding to $y=R/6$, we have $r=1$ and the integral (\ref{C213}) can be computed in closed form and the corresponding result reads
\begin{equation}
\label{C214}
\mathcal{G}_{A}^{\sigma\sigma\sigma}(x,R/6) = \frac{\textrm{erf}^{\, 3}(\eta)}{3} \, .
\end{equation}

\begin{figure}[htbp]
\centering
\includegraphics[width=105mm]{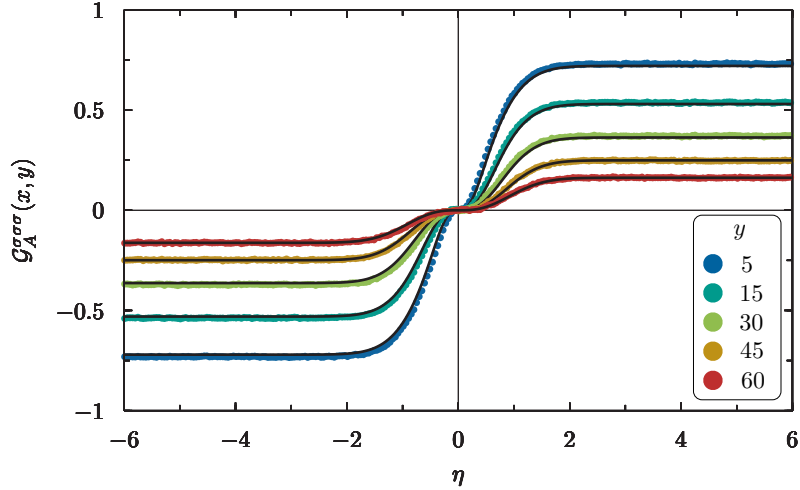}
\caption{The correlation function $\mathcal{G}_{A}^{\sigma\sigma\sigma}(x,y)$ for $T=2$ and $R=201$. Data points are denoted with symbols and correspond to the values of $y$ collected in the inset. Solid black curves correspond to the analytic result (\ref{C213}).}
\label{fig_ga}
\end{figure}

In Fig.~\ref{fig_ga}, we compare the numerical data of MC simulations obtained for $R=201$ and $T=2$ with the analytic result (\ref{C213}). A remarkable agreement is observed for a wide spectrum of $y$ ranging from $y=5$ ($\tau \approx 0.05$) to $y=60$ ($\tau \approx 0.60$). The horizontal asymptotes in Fig.~\ref{fig_ga} are given by (\ref{3024}), meaning that the clustering property (\ref{C211}) is confirmed by the simulations.

\begin{figure}[htbp]
\centering
\includegraphics[width=105mm]{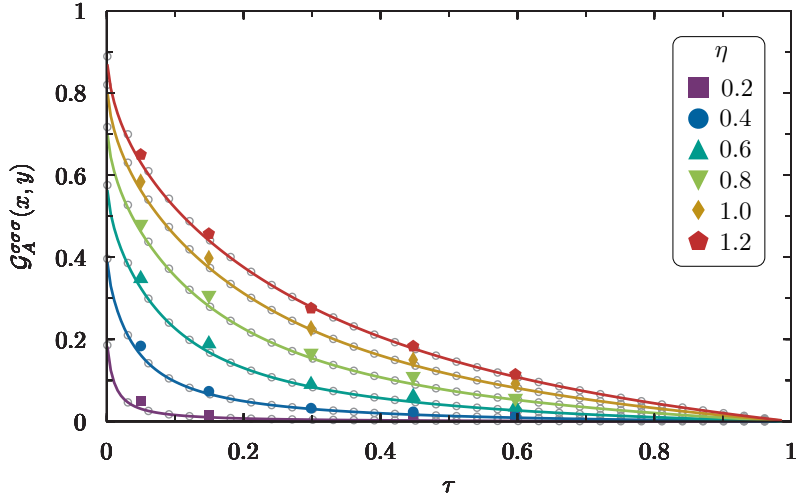}
\caption{The correlation function $\mathcal{G}_{A}^{\sigma\sigma\sigma}(x,y)$ as function of $\tau=2y/R$ for the values of $\eta=x/\lambda$ indicated in the inset. Numerical results are indicated with colored data points, the analytic result (\ref{C213}) is shown with solid lines with the same color code of numerical data. Empty gray circles indicate the approximation obtained by truncating the B\"urmann series of the error function including the term proportional to $b_{3}$; see Appendix \ref{AppendixC} for further details.}
\label{fig_galr}
\end{figure}

The occurrence of long range interfacial correlations can be tested in an explicit fashion by examining the decay of correlations upon increasing $y$ for fixed $x$. In Fig.~\ref{fig_galr}, we show the correlation function $\mathcal{G}_{A}^{\sigma\sigma\sigma}(x,y)$ as function of $\tau$ for several values of $\eta$.
In order to appreciate the long-range character exhibited by correlations along the interface, we compare the numerical results with the small-$\tau$ asymptotic expansion of the correlation function $\mathcal{G}_{A}^{\sigma\sigma\sigma}(x,y)$. Such a task is better achieved by examining the integral representation provided by (\ref{C213}). A rather simple expression obtained in the regime in which $\tau$ is small and $\eta \gtrsim C \sqrt{\tau}$, with $C=O(1)$ (see Appendix \ref{AppendixC}), reads
\begin{equation}
\begin{aligned}
\label{21112020_1427}
\mathcal{G}_{A}^{\sigma\sigma\sigma}(x,y) & \approx \textrm{erf}(\eta) - \frac{2\sqrt{2}}{\pi} \frac{1}{r(\tau)} \, , \\
& = \textrm{erf}(\eta) - \frac{4}{\pi} \sqrt{\tau} \, . 
\end{aligned}
\end{equation}
The term proportional to $\sqrt{\tau}=\sqrt{2y/R}$ indicates the occurrence of long range correlations. A rather more elaborate asymptotic expansion is needed in order to encompass the full interfacial region, which includes also $\eta \rightarrow0$. A very accurate description is provided by the following series representation
\begin{equation}
\label{21112020_1431}
\mathcal{G}_{A}^{\sigma\sigma\sigma}(x,y) = \textrm{erf}(\eta) - \sum_{n=1}^{\infty} B_{n} \frac{\textrm{erf}(\sqrt{1+nr^{2}}\eta)}{\sqrt{1+nr^{2}}} \, , \qquad r=\sqrt{\frac{1-\tau}{2\tau}} \, ,
\end{equation}
which we derive in Appendix \ref{AppendixC}. The coefficients $B_{n}$ can be extracted in a systematic fashion from the B\"urmann series of the error function. The series representation (\ref{21112020_1431}) is so accurate that it is enough to truncate the B\"urmann series up to the third-order term in order to achieve a perfect superposition between the series representation and the analytic result (\ref{C213}); see the empty circles in Fig.~\ref{fig_galr}.

It has to be noticed that the correlation function $\mathcal{G}_{A}^{\sigma\sigma\sigma}(x,y)$ exhibits a cubic behavior in the closeness of $x=0$. This property is manifestly evident for $y=R/6$ thanks to (\ref{C214}). From the series representation (\ref{21112020_1431}), we can actually appreciate that such a feature is valid for any $y$. Furthermore, the vanishing of $\mathcal{G}_{A}^{\sigma\sigma\sigma}(x,y)$ for large values of $y$ follows by observing that $r \rightarrow 0$ in such a limit; thus, (\ref{C213}) tends to zero. It is also possible to verify such a limiting behavior by inspecting the alternative expression provided by (\ref{21112020_1431}). In this case one needs to take $r \rightarrow0$ and use the property $\sum_{n=1}^{\infty}B_{n}=1$ proved in Appendix \ref{AppendixC}.

\subsubsection*{Configuration B}
The analytic expression for the correlator is given by
\begin{equation}
\begin{aligned}
\label{0312}
\mathcal{G}_{B}^{\sigma\sigma\sigma}(x,y) & = - 2\textrm{erf}(\chi) + 16 S_{-}\left(\sqrt{2}\chi,\frac{\varrho}{\sqrt{1-\varrho^{2}}},\frac{1}{\varrho}\right) + 16 S_{-}\left(\sqrt{2}\chi,\sqrt{\frac{1-\varrho^{2}}{1+\varrho^{2}}},\frac{2\varrho}{1-\varrho^{2}}\right) \, \\
& = -1 + 8T\left(\sqrt{2}\chi,\sqrt{\frac{1-\varrho^{2}}{1+\varrho^{2}}}\right) + \frac{2}{\sqrt{\pi}} \int_{-\infty}^{0}\textrm{d}u \, \textrm{e}^{-u^{2}} \textrm{erf}^{2}\left( \frac{\chi-\varrho u}{\sqrt{1-\varrho^{2}}} \right) \, ,
\end{aligned}
\end{equation}
where $S_{-}$ is the function defined by (\ref{B09}).

The expression in the upper line follows from the relationship between the cumulative distribution $\Phi_{3}$ and the functions $T$ and $S$. The expression in the second line can be obtained in a more direct route by carrying out the integrals with respect to $x_{1}$ and $x_{3}$ in (\ref{4005}), while the integral on $x_{2}$ remains in the implicit form shown in (\ref{0312}). Reflection symmetry implies that $\mathcal{G}_{B}^{\sigma\sigma\sigma}(x,y)$ is odd with respect to $x$ for fixed $y$. The vanishing of (\ref{0312}) for $x=0$ is consistent with the general properties discussed in Sec.~\ref{sec42}; see e.g. (\ref{0310}). Although the second expression may be advantageous for numerical implementations, the above symmetries are not manifest. On the other hand, the expression in the first line shows the required symmetries explicitly.

\begin{figure}[htbp]
\centering
\includegraphics[width=105mm]{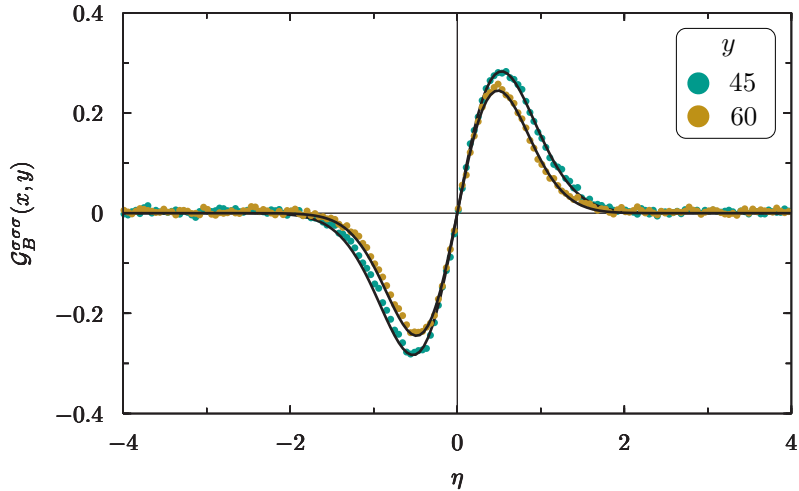}
\caption{The correlation function $\mathcal{G}_{B}^{\sigma\sigma\sigma}(x,y)$ for $T=2$ and $R=201$. Data points are indicated with dots and the values of $y$ are indicated in the inset. Solid black curves correspond to the analytic result (\ref{0312}).}
\label{fig_gb}
\end{figure}

Upon taking the limit $|x| \rightarrow \infty$, we find the clustering property $\langle \sigma(x,y) \sigma (0,0) \sigma(x,-y) \rangle_{-+} \rightarrow \langle \sigma (0,0) \rangle_{-+}  \langle \sigma(x,y) \sigma(x,-y) \rangle_{-+} =0$, thus (\ref{0312}) tends to zero for large $|x|$. The agreement between the analytic result (\ref{0312}) and MC simulations is shown in Fig.~\ref{fig_gb}.

\subsubsection*{Configuration C}
The analytic expression for the correlator is given by
\begin{equation}
\begin{aligned}
\label{4180}
\mathcal{G}_{C}^{\sigma\sigma\sigma}(x,y) & = \textrm{erf}(\eta) + 2\textrm{erf}(\chi) - 16 S_{-}\left( \sqrt{2}\chi, \frac{\varrho\sqrt{1-\varrho^{2}}}{1+\varrho^{2}},\frac{1}{\varrho} \right) - 16 S_{-}\left( \sqrt{2}\eta, \frac{\sqrt{1-\varrho^{2}}}{2\varrho},1 \right) \, .
\end{aligned}
\end{equation}
The correlation function $\mathcal{G}_{C}^{\sigma\sigma\sigma}(x,y)$ is odd with respect to $x \rightarrow -x$. This symmetry is manifest in (\ref{4180}). Thanks to the identities satisfied by the function $S_{-}$ (see Appendix \ref{appendixB}), we can establish the clustering property $\langle \sigma(x,y) \sigma (x,0) \sigma(x,-y) \rangle_{-+} \rightarrow \pm M^{3}$ for $x\rightarrow \pm \infty$; or equivalently $\mathcal{G}_{C}^{\sigma\sigma\sigma}(x \rightarrow \pm \infty,y) \rightarrow \pm 1$. Of course, the above writing refers only to the degrees of freedom coupled to the interface and not to the bulk three-point function, as already stressed.

\begin{figure}[htbp]
\centering
\includegraphics[width=105mm]{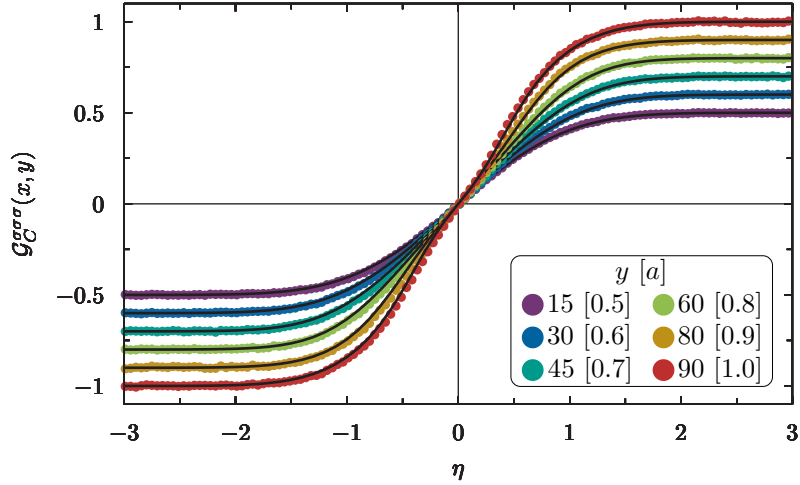}
\caption{The correlation function $\mathcal{G}_{C}^{\sigma\sigma\sigma}(x,y) \times a(y)$ for $T=2$ and $R=201$. Circles refer to numerical data and solid black curves are computed from the analytic result (\ref{4180}). The inset indicates the values of $y$ and the multiplicative factor $a(y)$ used in order to displace the curves.}
\label{fig_gc}
\end{figure}

The excellent agreement between theory and numerics is confirmed in Fig.~\ref{fig_gc}. Since the $y$-dependence of (\ref{4180}) turns out to be rather weak, curves corresponding to different values of $y$ result very close to each other. In order to better visualize all the data sets, both the numerical and analytical results in Fig.~\ref{fig_gc} are multiplied by a coefficient $a(y)$ which takes different values for those values of $y$ sampled in Fig.~\ref{fig_gc}. For each data set, we indicate the coefficient $a(y)$ into a square bracket.

\newpage
\section{Conclusions}
\label{sec5}
In this paper we have tested several predictions of the exact theory of phase separation against high-quality Monte Carlo simulations for the Ising model with boundary conditions enforcing an interface on the strip. An excellent agreement between theory and numerics is observed for order parameter correlation functions at the leading order in finite size corrections. For the magnetization profile, we have isolated the leading finite-size correction obtained from the numerical data and found a good agreement when tested against the theoretical prediction. We have shown how to extract the passage probability density for off-critical interfaces directly from numerical simulations. Albeit the Gaussian nature of off-critical interfacial fluctuation is a well-established result\footnote{See \cite{Gallavotti_72,GI,CIV} for rigorous reasults and \cite{fisher_walks_1984} for heuristic arguments.}, the methodology employed in this paper can be applied to the study of those universality classes in certain geometries for which exact results are not yet available. As a specific example in which exact results are available, the aforementioned technique has been recently employed in the study of correlation functions in the presence of a wall \cite{ST_droplet}. 

Lastly, the long-range character of interfacial correlations has been established by means of explicit calculations of both two- and three-point correlation functions of the order parameter field for several spatial arrangements of spin fields. The numerical result are again in excellent agreement with the theory in absence of adjustable parameters. Although in this paper we have considered $n=1,2$ and $3$ point-correlation functions for the Ising model, computer simulation studies and closed form expressions can be obtained also for four-point correlation functions. These results will appear in a companion paper \cite{ST_fourpoint}.

\section*{Acknowledgements}
A. S. is grateful to Gesualdo Delfino for his valuable comments and to Douglas B. Abraham for many interesting discussions and for collaborations on closely related topics. A.~S and A.~T. acknowledge Oleg Vasilyev for precious hints about numerical algorithms. A.~S.~acknowledges the Galileo Galilei Institute for Theoretical Physics (Arcetri, Florence) for hospitality received in the early stages of this work during the event \emph{``SFT 2019: Lectures on Statistical Field Theories''}.

\appendix

\section{Computational toolbox}
\label{appendixB}
The cumulative distribution functions for the Gaussian bivariate and trivariate distributions, respectively $\Phi_{2}$ and $\Phi_{3}$, can be expressed in terms of a certain class of special functions known as Owen's $T$ and Steck's $S$ functions. For the sake of convenience, we report the most relevant mathematical properties of the functions $T$ and $S$ which are useful in the manipulations of the correlation functions (\ref{3023}) and (\ref{11022021_1325}). We refer the interested reader to \cite{Owen1980} for a thorough exposition on cumulative distribution functions of Gaussian distributions.

\subsection{Owen's $T$ and Steck's $S$ functions}
We begin by recalling the expression of the trivariate normal distribution
\begin{equation}
\label{B01_00}
\Pi_{3}(u_{1},u_{2},u_{3} \vert \rho_{12}, \rho_{13}, \rho_{23}) = \frac{ \textrm{e}^{-v/\Delta} }{ (2\pi)^{3/2} \sqrt{\Delta} } \, ,
\end{equation}
with
\begin{equation}
\Delta = 1 - \rho_{12}^{2} - \rho_{13}^{2} - \rho_{23}^{2} + 2 \rho_{12} \rho_{13} \rho_{23} \, ,
\end{equation}
and
\begin{equation}
\begin{aligned}
\label{}
v & = (1-\rho_{23}^{2}) u_{1}^{2} + (1-\rho_{13}^{2}) u_{2}^{2} + (1-\rho_{12}^{2}) u_{3}^{2} - 2 (\rho_{12}-\rho_{13}\rho_{23}) u_{1} u_{2} \\
& - 2 (\rho_{13}-\rho_{12}\rho_{23}) u_{1} u_{3} - 2 (\rho_{23}-\rho_{12}\rho_{13}) u_{2} u_{3} \, .
\end{aligned}
\end{equation}
Being (\ref{B01_00}) a standardized distribution, one has $\mathbb{E}[ u_{j} ]=0$ and $\mathbb{E}[ u_{j}^{2} ]=1$ for $j=1,2,3$, while $\mathbb{E}[ u_{i}u_{j} ]=\rho_{ij}$ for $i \neq j$.

Owen $T$-function is defined by the integral
\begin{equation}
\label{B01}
T(h,a) = \frac{1}{2\pi} \int_{0}^{a}\textrm{d}x \, \frac{ \textrm{e}^{-(1+x^{2})h^{2}/2} }{ 1+x^{2} } \, .
\end{equation}
The function $T(h,a)$ satisfies the symmetry properties $T(h,a) = T(-h,a) = - T(h,-a)$ and $T(h,0)=T(\pm\infty,0)=0$. For special values of its arguments, $T(h,a)$ reduces to
\begin{align}
\label{Owen_special_1}
T(0,a) & = \frac{1}{2\pi} \tan^{-1}a \, , \\
\label{Owen_special_2}
T(\sqrt{2}h,1) & = \frac{1-\textrm{erf}^{2}(h)}{8} = \frac{1}{8} \textrm{erfc}(h) \textrm{erfc}(-h) \, , \\
\label{Owen_special_3}
T(\sqrt{2}h,\pm \infty) & = \pm \frac{1}{4} \textrm{erfc}(|h|) \, ,
\end{align}
where $\textrm{erf}(x)=(2/\sqrt{\pi})\int_{0}^{x} \textrm{d}u\, \exp(-u^{2})$ is the error function and $\textrm{erfc}(x)=1-\textrm{erf}(x)$ is the complementary error function.

Steck $S$-function can be defined by the integral
\begin{equation}
\label{B02}
S(h,a,b) = \frac{1}{\sqrt{2\pi}}\int_{-\infty}^{h}\textrm{d}x \, T(ax,b) \textrm{e}^{-x^{2}/2} \, .
\end{equation}
There exist a number of equivalent integral representation of $S(h,a,b)$ which are convenient for numerical implementations, for instance
\begin{equation}
\label{B03}
S(h,a,b) = \frac{1}{2\pi} \int_{0}^{b}\textrm{d}x \, \frac{ \Phi_{1}(h\sqrt{1+a^{2}+a^{2}x^{2}}) }{ (1+x^{2})\sqrt{1+a^{2}+a^{2}x^{2}} } \, ,
\end{equation}
with $\Phi_{1}(x)$ the cumulative distribution of the univariate normal Gaussian; see (\ref{3020}) and (\ref{3020_a}). The function $S(h,a,b)$ satisfies the following properties:
\begin{eqnarray}
\label{B05}
S(-\infty,a,b) & = & S(h,\pm\infty,b) = 0 \\
\label{B06}
S(+\infty,a,b) & = & 2S(0,a,b) = \frac{1}{2\pi}\tan^{-1}\frac{b}{\sqrt{1+a^{2}+a^{2}b^{2}}} \\
\label{B07}
S(-h,a,b) & = & S(+\infty,a,b) - S(h,a,b) \\
\label{B08}
S(h,-a,b) & = & S(h,a,b) = -S(h,a,-b) \, .
\end{eqnarray}
For the purpose of further elaborations, we decompose $S(h,a,b)$ in terms of its even and odd parts with respect to the variable $h$, i.e. $S(h,a,b) = S_{+}(h,a,b) + S_{-}(h,a,b)$, with the even and odd parts given by $S_{\pm}(h,a,b) = (1/2) (S(h,a,b) \pm S(-h,a,b)$. Thanks to (\ref{B07}) and (\ref{B06}), it follows that $S_{+}(h,a,b) = S(0,a,b)$; thus, the even part does not depend on $h$. The odd part instead can be written as follows
\begin{equation}
\label{B09}
S_{-}(h,a,b) = \frac{1}{\sqrt{2\pi}}\int_{0}^{h}\textrm{d}x \, T(ax,b) \textrm{e}^{-x^{2}/2} \, ,
\end{equation}
which is manifestly odd with respect to $h$. It thus follows that $S_{-}(0 ,a,b)=0$ and 
\begin{equation}
\label{B11}
S_{-}(\pm \infty ,a,b) = \pm S(0,a,b) \, ,
\end{equation}
while the symmetries with respect to $a$ and $b$ are the same of $S(h,a,b)$. We also quote the following identity
\begin{equation}
\label{04042021_1150}
\partial_{h} S_{-}(\sqrt{2}h,a,1) = \frac{1-\textrm{erf}^{2}(ah)}{8\sqrt{\pi}} \textrm{e}^{-h^{2}} \, ,
\end{equation}
which is useful in order to prove (\ref{C213}).

\subsection{Three-point correlation function in configurations A, B, and C}
Establishing (\ref{C213}), (\ref{0312}), and (\ref{4180}) is a straightforward (although rather tedious) calculation which can be done by using Eq. (3.7) of \cite{Owen1980}, the latter expresses $\Phi_{3}$ in terms of $T$ and $S$. The symmetric arrangements $A$, $B$ and $C$ are realized by special values of the correlation coefficients ($\rho_{12}=\rho_{23}=\sqrt{\rho_{13}}$), the latter are responsible for drastic simplifications in Eq. (3.7) of \cite{Owen1980}.

Let us begin with configuration $A$. By expressing $\Phi_{3}$ in terms of $T$ and $S$ functions \cite{Owen1980}, the general result (\ref{4006}) for the spin fields in configuration $A$ gives
\begin{equation}
\begin{aligned}
\label{0311}
\mathcal{G}_{A}^{\sigma\sigma\sigma}(x,y) & = \langle \sigma(0,y) \sigma(x,0) \sigma(0,-y) \rangle_{-+}/M^{3} \, , \\
& = 1+\textrm{erf}(\eta) - \frac{2}{\pi}\sin^{-1}\varrho^{2} - 16 S \left(\sqrt{2}\eta,\frac{\varrho}{\sqrt{1-\varrho^{2}}},1 \right) \, .
\end{aligned}
\end{equation}
By bringing in the even and odd parts of $S(h,a,b)$ and using the following relationship
\begin{equation}
\label{ }
S \left(0,\frac{\varrho}{\sqrt{1-\varrho^{2}}},1 \right) =  \frac{1}{16} - \frac{1}{8\pi}\sin^{-1}\varrho^{2} \, ,
\end{equation}
the correlation function $\mathcal{G}_{A}^{\sigma\sigma\sigma}(x,y)$ becomes
\begin{equation}
\begin{aligned}
\label{4010}
\mathcal{G}_{A}^{\sigma\sigma\sigma}(x,y) & = \textrm{erf}(\eta) - 16 S_{-}\left(\sqrt{2}\eta,r,1 \right) \, , \qquad r=\frac{\varrho}{\sqrt{1-\varrho^{2}}} \, ,
\end{aligned}
\end{equation}
where $S_{-}$ is the function defined in (\ref{B09}). Quite interestingly, the first derivative of $\mathcal{G}_{A}^{\sigma\sigma\sigma}(x,y)$ with respect to $\eta$ admits a remarkably simple expression. Thanks to the identity (\ref{04042021_1150})
\begin{equation}
\label{C212}
\partial_{\eta} \mathcal{G}_{A}^{\sigma\sigma\sigma}(x,y) = \frac{2}{\sqrt{\pi}} \textrm{erf}^{\, 2}(r\eta) \textrm{e}^{-\eta^{2}} \, ,
\end{equation}
therefore, integrating with respect to $\eta$ and using the known value in the origin, we find the integral representation (\ref{C213}), which is equivalent to (\ref{4010}).

For the configuration $B$ an analogous calculation leads us to
\begin{equation}
\begin{aligned}
\mathcal{G}_{B}^{\sigma\sigma\sigma}(x,y) & = \langle \sigma(x,y) \sigma(0,0) \sigma(x,-y) \rangle_{-+}/M^{3} \, , \\
& = 16 S \left(\sqrt{2}\chi,\frac{\varrho}{\sqrt{1-\varrho^{2}}},\frac{1}{\varrho}\right) + 16 S \left(\sqrt{2}\chi,\sqrt{\frac{1-\varrho^{2}}{1+\varrho^{2}}},\frac{2\varrho}{1-\varrho^{2}}\right) - 2\textrm{erf}(\chi)-2 \, .
\end{aligned}
\end{equation}
We observe the following property
\begin{equation}
\label{ }
S \left(0,\frac{\varrho}{\sqrt{1-\varrho^{2}}},\frac{1}{\varrho}\right) + S \left(0,\sqrt{\frac{1-\varrho^{2}}{1+\varrho^{2}}},\frac{2\varrho}{1-\varrho^{2}}\right) = \frac{1}{8} \, .
\end{equation}
By using the decomposition of $S$ into even and odd parts, we find a vanishing even part. The result is thus the odd function given in (\ref{0312}).

Lastly, we consider the configuration $C$. By following the same guidelines outlined for cases $A$ and $B$, we find
\begin{equation}
\begin{aligned}
\label{0313}
\mathcal{G}_{C}^{\sigma\sigma\sigma}(x,y) & = \langle \sigma(x,y) \sigma(x,0) \sigma(x,-y) \rangle_{-+}/M^{3} \, , \\
& = 2 + \textrm{erf}(\eta) + 2\textrm{erf}(\chi) - 16 S\left( \sqrt{2}\chi, \frac{\varrho\sqrt{1-\varrho^{2}}}{1+\varrho^{2}},\frac{1}{\varrho} \right) - 16 S\left( \sqrt{2}\eta, \frac{\sqrt{1-\varrho^{2}}}{2\varrho},1 \right) \, .
\end{aligned}
\end{equation}
Although it could be not obvious from (\ref{0313}), the profile $\mathcal{G}_{C}^{\sigma\sigma\sigma}(x,y)$ is an odd function of $x$ which interpolates between $\pm 1$, the latter are the asymptotic values reached for $x \rightarrow \pm \infty$. In order to exhibit such a symmetry in a manifest fashion, it is convenient to express $S$ in terms of its even and odd parts. Thanks to the property
\begin{equation}
\label{ }
S\left( 0, \frac{\varrho\sqrt{1-\varrho^{2}}}{1+\varrho^{2}},\frac{1}{\varrho} \right) + S\left( 0, \frac{\sqrt{1-\varrho^{2}}}{2\varrho},1 \right) = \frac{1}{8}
\end{equation}
it is immediate to find (\ref{4180}).

\section{B\"urmann series representations}
\label{AppendixC}
The correlation function for spins in configuration $A$ can be written in the following form
\begin{equation}
\label{21112020_1456}
\mathcal{G}_{A}^{\sigma\sigma\sigma}(x,y) = \textrm{erf}(\eta) - \frac{2}{\sqrt{\pi}} \int_{0}^{\eta}\textrm{d}u \, \left( 1- \textrm{erf}^{2}(ru) \right) \textrm{e}^{-u^{2}} \, ,
\end{equation}
where $r=r(\tau)$ is given by (\ref{21112020_1431}). Thanks to B\"urmann theorem \cite{WW,SP_2014}, the error function can be expressed in terms of the following series
\begin{align}
\label{21112020_1457}
\textrm{erf}(x) & = \frac{2}{\sqrt{\pi}} \textrm{sign}(x) \sqrt{1-\textrm{e}^{-x^{2}}} \biggl[ 1 + \sum_{n=1}^{\infty} b_{n} \left( 1-\textrm{e}^{-x^{2}}\right)^{n} \biggr] \, ,
\end{align}
which converges rapidly to the error function for any real value of $x$. The first B\"urmann coefficients are given by $b_{1}=-1/12$, $b_{2}=-7/480$, $b_{3}=-5/896$, $b_{4}=-787/276480$. We mention also the following alternative representation
\begin{align}
\label{22112020_0955}
\textrm{erf}(x) & = \frac{2}{\sqrt{\pi}} \textrm{sign}(x) \sqrt{1-\textrm{e}^{-x^{2}}} \sum_{n=0}^{\infty} c_{n} \textrm{e}^{-nx^{2}} \, , \qquad c_{0}=\sqrt{\pi}/2 \, ,
\end{align}
which is particularly suited for numerical evaluations. The truncation of the aforementioned series to the first two exponentials with coefficients $c_{1}=c_{1}^{\rm num}=31/200$ and $c_{2}=c_{2}^{\rm num}=-341/8000$ provides a very accurate representation of the error function for numerical purposes \cite{SP_2014}. Thanks to the binomial theorem we can pass from the series representation (\ref{21112020_1457}) to (\ref{22112020_0955}) and identify the exact relationship between coefficients $b_{n}$ and $c_{n}$
\begin{equation}
\begin{aligned}
\label{21112020_1656}
c_{0} & = \sum_{n=0}^{\infty} b_{n} \, , \\
c_{1} & = - \sum_{n=1}^{\infty} n b_{n} \, , \\
c_{2} & = \frac{1}{2} \sum_{n=2}^{\infty} n(n-1) b_{n} \, , \\
c_{3} & = - \frac{1}{6} \sum_{n=3}^{\infty} n(n-1)(n-2) b_{n} \, , \\
\end{aligned}
\end{equation}
or, by induction, for $n \geqslant 0$ we can desume
\begin{equation}
\label{21112020_1657}
c_{n} = (-1)^{n} \sum_{j=n}^{\infty} \binom{j}{n} b_{j} \, , \qquad b_{0}=1 \, .
\end{equation}

Coming back to the evaluation of (\ref{21112020_1456}), the form of both series (\ref{21112020_1457}) and (\ref{22112020_0955}) is not particularly adapt for the calculation of the integral. We thus consider the following rearrangement
\begin{equation}
\begin{aligned}
\label{21112020_1538}
1-\textrm{erf}^{2}(x) & =  \sum_{n=1}^{\infty} B_{n} \textrm{e}^{-nx^{2}} \, .
\end{aligned}
\end{equation}
In order to find the relationship between the coefficients $B_{n}$ and the coefficients $b_{n}$, we equate the series (\ref{21112020_1538}) and the B\"urmann series (\ref{21112020_1457}), and identify term by term. The procedure is actually facilitated by working with the variable $z=1-\textrm{e}^{-x^{2}}$. The identification thus implies
\begin{equation}
\label{ }
1-\frac{4}{\pi} z \left( \sum_{n=0}^{\infty} b_{n} z^{n} \right)^{2} = \sum_{n=1}^{\infty} B_{n} (1-z)^{n} \, .
\end{equation}
By plugging $z=1$ the right hand side vanishes and also the left one does because $\sum_{n=0}^{\infty} b_{n}=\sqrt{\pi}/2$, with $b_{0}=1$; this last identity follows by taking $x\rightarrow + \infty$ in (\ref{21112020_1457}). Subsequent $B_{n}(\{b_{n}\})$ can be extracted by further Taylor expanding around $z=1$. On the other hand, $b_{n}(\{B_{n}\})$ can be determined by the Taylor expansion around $z=0$. Another way of extracting $B_{n}(\{b_{n}\})$ is by matching (\ref{21112020_1538}) with the square of (\ref{21112020_1457}) term by term. A direct evaluation yields
\begin{equation}
\begin{aligned}
\label{21112020_1827}
B_{1} & = 1-2c_{1} \\
B_{2} & = 2c_{1} -c_{1}^{2} -2c_{2} \\
B_{3} & = c_{1}^{2} + 2c_{2} -2c_{1}c_{2} -2c_{3} \\
B_{4} & = 2 c_{1} c_{2} + 2 c_{3} - 2 c_{1} c_{3} - c_{2}^2 
\end{aligned}
\end{equation}
with $c_{n}(\{b_{n}\})$ given by (\ref{21112020_1657}). Once we have extracted the coefficients $B_{n}$, by plugging (\ref{21112020_1538}) into (\ref{21112020_1456}), a simple calculation entails
\begin{equation}
\label{21112020_1544}
\mathcal{G}_{A}^{\sigma\sigma\sigma}(x,y) = \textrm{erf}(\eta) - \sum_{n=1}^{\infty} B_{n} \frac{\textrm{erf}(\sqrt{1+nr^{2}}\eta)}{\sqrt{1+nr^{2}}} \, ,
\end{equation}
which is the result (\ref{21112020_1431}) given in the main body of the paper.

We further comment on two limiting behaviors. Notice that by setting $x=0$ in (\ref{21112020_1538}) we obtain the condition\footnote{Series of the form $\sum_{n=1}^{\infty}n^{s}B_{n}$ with $s\in\mathbb{N}$ can be evaluated by differentiating (\ref{21112020_1538}) with respect to $x$ at $x=0$. Series of the form $\sum_{n=1}^{\infty} n^{-s-1/2} B_{n}$ and $\sum_{n=1}^{\infty} n^{-s} B_{n}$ instead follow by taking moments of (\ref{21112020_1538}) for $x \in (-\infty,+\infty)$ or $x \in (0,+\infty)$.} $\sum_{n=1}^{\infty}B_{n}=1$. Then, in the limit of large vertical separation between spin fields the correlation function vanishes. Such a feature can be easily established by observing that for $\tau \rightarrow 1$ we have $r \rightarrow 0$ and both terms in (\ref{21112020_1544}) become identical by virtue of the property $\sum_{n=1}^{\infty}B_{n}=1$. Secondly, the plot of $\mathcal{G}_{A}^{\sigma\sigma\sigma}(x,y)$ versus $x$ is characterized by a vanishing slope for $x=0$. This feature, which is evident from (\ref{21112020_1456}), follows from (\ref{21112020_1544}) thanks to the aforementioned property that the sum of the $B_{n}$'s is one.

Let us consider now the limit of small vertical separation between spin fields, so $\tau \rightarrow0$ and $r \rightarrow + \infty$. If the rescaled abscissa of the two spin fields is $\eta>0$, then the smallest argument in the error functions appearing in (\ref{21112020_1544}) is $\sqrt{1+r^{2}} \eta \approx \eta/\sqrt{2\tau}$. Provided $\eta/\sqrt{2\tau}$ is large in the sense that $\textrm{erf}(\eta/\sqrt{2\tau})>1-\epsilon$, with $\epsilon>0$ a small parameter, then it follows that all error functions in (\ref{21112020_1544}) are bounded from below by $1-\epsilon$. The expression (\ref{21112020_1544}) can be bounded with
\begin{equation}
\label{}
\mathcal{G}_{A}^{\sigma\sigma\sigma}(x,y) < \textrm{erf}(\eta) - (1-\epsilon) \sum_{n=1}^{\infty} \frac{B_{n}}{\sqrt{1+n r^{2}}} \, .
\end{equation}
In the limit of large $r$ and small $\epsilon$, which is the one we are interested in, the above can be approximated as follows
\begin{equation}
\label{11042021_1437}
\mathcal{G}_{A}^{\sigma\sigma\sigma}(x,y) \approx \textrm{erf}(\eta) - \frac{1}{r} \sum_{n=1}^{\infty} \frac{B_{n}}{\sqrt{n}} \, .
\end{equation}
The series appearing in the above can be evaluated in closed form and reads as follows
\begin{equation}
\label{11042021_1436}
\sum_{n=1}^{\infty} \frac{B_{n}}{\sqrt{n}} = \frac{2\sqrt{2}}{\pi} \, ,
\end{equation}
the above identity follows straightforwardly upon integrating (\ref{21112020_1538}) with respect to $x$ from $-\infty$ to $+\infty$. By inserting (\ref{11042021_1436}) into (\ref{11042021_1437}), we obtain (\ref{21112020_1427}). It has to be emphasized that the condition $\textrm{erf}(\eta/\sqrt{2\tau}) > 1-\epsilon$ translates into
\begin{equation}
\label{11042021_1438}
\eta > C(\epsilon) \sqrt{\tau} \, ,
\end{equation}
where $C(\epsilon)=\textrm{erf}^{-1}(1-\epsilon)$, where $\textrm{erf}^{-1}$ is the inverse of the error function. For instance, if we set $\epsilon=0.01$, corresponding to $\textrm{erf}(\eta/\sqrt{2\tau})>0.99$, then we find $C(\epsilon) \approx 1.821$. Thus, for reasonably small values of $\epsilon$, the constant $C(\epsilon)$ is of order $1$. The condition (\ref{11042021_1438}) sets the domain of validity of the approximation (\ref{11042021_1437}).

Checking the clustering to the asymptotic value for $\eta \rightarrow +\infty$ requires additional efforts. From (\ref{21112020_1544}), we find
\begin{equation}
\label{ }
\lim_{\eta \rightarrow +\infty} \mathcal{G}_{A}^{\sigma\sigma\sigma}(x,y) = 1 - \sum_{n=1}^{\infty} \frac{B_{n} }{\sqrt{1+nr^{2}}} \, ,
\end{equation}
thanks to the identity
\begin{equation}
\label{30112020_1129}
\sum_{n=1}^{\infty} \frac{B_{n} }{\sqrt{1+nr^{2}}} = 2 - \frac{4}{\pi}\tan^{-1}\sqrt{1+2r^{2}} \, ,
\end{equation}
we obtain
\begin{equation}
\label{ }
\lim_{\eta \rightarrow +\infty} \mathcal{G}_{A}^{\sigma\sigma\sigma}(x,y) = - 1 + \frac{4}{\pi}\tan^{-1}\sqrt{1+2r^{2}} \, ,
\end{equation}
which coincides with the clustering relation (\ref{13042021_2331}).

For the sake of completeness, we observe how the identity (\ref{30112020_1129}) contains the property $\sum_{n=1}^{\infty}B_{n}=1$ as a special case. The above can be used in order to evaluate series of the form $\sum_{n=1}^{\infty} n^{k} B_{n}$ with $k \in \mathbb{N}$ by expanding in powers series of $r$ both sides of (\ref{30112020_1129}) and equating order by order in powers of $r$. This procedure is actually analogous to the recursive scheme which follows by evaluating derivatives with respect to $x$ at $x=0$ for the series representation (\ref{21112020_1538}).

\section{Mixed three-point correlation functions}
\label{AppendixD}
\begin{figure*}[htbp]
\centering
        \begin{subfigure}[b]{0.23\textwidth}
            \centering
\begin{tikzpicture}[thick, line cap=round, >=latex, scale=0.17]
\tikzset{fontscale/.style = {font=\relsize{#1}}}
\draw[thin, dashed, -] (-10, 0) -- (11, 0) node[below] {};
\draw[thin, dashed, -] (0, -7) -- (0, 7) node[left] {};
\draw[very thick, red, -] (0, 7) -- (10, 7) node[] {};
\draw[very thick, red, -] (0, -7) -- (10, -7) node[] {};
\draw[very thick, blue, -] (-10, 7) -- (0, 7) node[] {};
\draw[very thick, blue, -] (-10, -7) -- (0, -7) node[] {};
\draw[thin, fill=green!30] (0, 4.9) circle (10pt) node[right] { $(0,y)$ };;
\draw[thin, fill=myyellow!80] (4.4, 0) circle (10pt) node[below] { $(x,0)$ };;
\draw[thin, fill=green!30] (0, -4.9) circle (10pt) node[right] { $(0,-y)$ };;
\draw[thin, fill=white] (-7, 7) circle (0pt) node[above] {${\color{blue}{-}}$};;
\draw[thin, fill=white] (-7, -7) circle (0pt) node[below] {${\color{blue}{-}}$};;
\draw[thin, fill=white] (7, 7) circle (0pt) node[above] {${\color{red}{+}}$};;
\draw[thin, fill=white] (7, -7) circle (0pt) node[below] {${\color{red}{+}}$};;
\end{tikzpicture}
            \caption[]%
            {{\small $\mathcal{G}_{A}^{\sigma\varepsilon\sigma}(x,y)$}}    
            \label{fig_sesa}
        \end{subfigure}
        \begin{subfigure}[b]{0.26\textwidth}
            \centering
\begin{tikzpicture}[thick, line cap=round, >=latex, scale=0.17]
\tikzset{fontscale/.style = {font=\relsize{#1}}}
\draw[thin, dashed, -] (-10, 0) -- (11, 0) node[below] {};
\draw[thin, dashed, -] (0, -7) -- (0, 7) node[left] {};
\draw[very thick, red, -] (0, 7) -- (10, 7) node[] {};
\draw[very thick, red, -] (0, -7) -- (10, -7) node[] {};
\draw[very thick, blue, -] (-10, 7) -- (0, 7) node[] {};
\draw[very thick, blue, -] (-10, -7) -- (0, -7) node[] {};
\draw[thin, fill=green!30] (9.2, 4.9) circle (10pt) node[left] { $(x,y)$ };;
\draw[thin, fill=myyellow!80] (0, 0) circle (10pt) node[below] { $(0,0)$ };;
\draw[thin, fill=green!30] (9.2, -4.9) circle (10pt) node[left] { $(x,-y)$ };;
\draw[thin, fill=white] (-7, 7) circle (0pt) node[above] {${\color{blue}{-}}$};;
\draw[thin, fill=white] (-7, -7) circle (0pt) node[below] {${\color{blue}{-}}$};;
\draw[thin, fill=white] (7, 7) circle (0pt) node[above] {${\color{red}{+}}$};;
\draw[thin, fill=white] (7, -7) circle (0pt) node[below] {${\color{red}{+}}$};;
\end{tikzpicture}
            \caption[]%
            {{\small $\mathcal{G}_{B}^{\sigma\varepsilon\sigma}(x,y)$}}    
            \label{fig_sesb}
        \end{subfigure}
\hfill
        \begin{subfigure}[b]{0.23\textwidth}
            \centering
\begin{tikzpicture}[thick, line cap=round, >=latex, scale=0.17]
\tikzset{fontscale/.style = {font=\relsize{#1}}}
\draw[thin, dashed, -] (-10, 0) -- (11, 0) node[below] {};
\draw[thin, dashed, -] (0, -7) -- (0, 7) node[left] {};
\draw[very thick, red, -] (0, 7) -- (10, 7) node[] {};
\draw[very thick, red, -] (0, -7) -- (10, -7) node[] {};
\draw[very thick, blue, -] (-10, 7) -- (0, 7) node[] {};
\draw[very thick, blue, -] (-10, -7) -- (0, -7) node[] {};
\draw[thin, fill=green!30] (8.0, 4.9) circle (10pt) node[left] { $(x,y)$ };;
\draw[thin, fill=myyellow!80] (8.0, 0) circle (10pt) node[below] { $(x,0)$ };;
\draw[thin, fill=green!30] (8.0, -4.9) circle (10pt) node[left] { $(x,-y)$ };;
\draw[thin, fill=white] (-7, 7) circle (0pt) node[above] {${\color{blue}{-}}$};;
\draw[thin, fill=white] (-7, -7) circle (0pt) node[below] {${\color{blue}{-}}$};;
\draw[thin, fill=white] (7, 7) circle (0pt) node[above] {${\color{red}{+}}$};;
\draw[thin, fill=white] (7, -7) circle (0pt) node[below] {${\color{red}{+}}$};;
\end{tikzpicture}
            \caption[]%
            {{\small $\mathcal{G}_{C}^{\sigma\varepsilon\sigma}(x,y)$}}    
            \label{fig_sesc}
        \end{subfigure}
\hfill
        \begin{subfigure}[b]{0.23\textwidth}
            \centering
\begin{tikzpicture}[thick, line cap=round, >=latex, scale=0.17]
\tikzset{fontscale/.style = {font=\relsize{#1}}}
\draw[thin, dashed, -] (-10, 0) -- (11, 0) node[below] {};
\draw[thin, dashed, -] (0, -7) -- (0, 7) node[left] {};
\draw[very thick, red, -] (0, 7) -- (10, 7) node[] {};
\draw[very thick, red, -] (0, -7) -- (10, -7) node[] {};
\draw[very thick, blue, -] (-10, 7) -- (0, 7) node[] {};
\draw[very thick, blue, -] (-10, -7) -- (0, -7) node[] {};
\draw[thin, fill=green!30] (9.2, 4.9) circle (10pt) node[left] { $(x,y)$ };;
\draw[thin, fill=myyellow!80] (0, 0) circle (10pt) node[below] { $(0,0)$ };;
\draw[thin, fill=green!30] (-9.2, -4.9) circle (10pt) node[right] { $(-x,-y)$ };;
\draw[thin, fill=white] (-7, 7) circle (0pt) node[above] {${\color{blue}{-}}$};;
\draw[thin, fill=white] (-7, -7) circle (0pt) node[below] {${\color{blue}{-}}$};;
\draw[thin, fill=white] (7, 7) circle (0pt) node[above] {${\color{red}{+}}$};;
\draw[thin, fill=white] (7, -7) circle (0pt) node[below] {${\color{red}{+}}$};;
\end{tikzpicture}
            \caption[]%
            {{\small $\mathcal{G}_{D}^{\sigma\varepsilon\sigma}(x,y)$}}    
            \label{fig_sesd}
        \end{subfigure}
\vskip\baselineskip
\caption[]
{\small Mixed three-point correlation functions involving two spin fields and one energy density field.}
\label{fig_ssemixed}
\end{figure*}
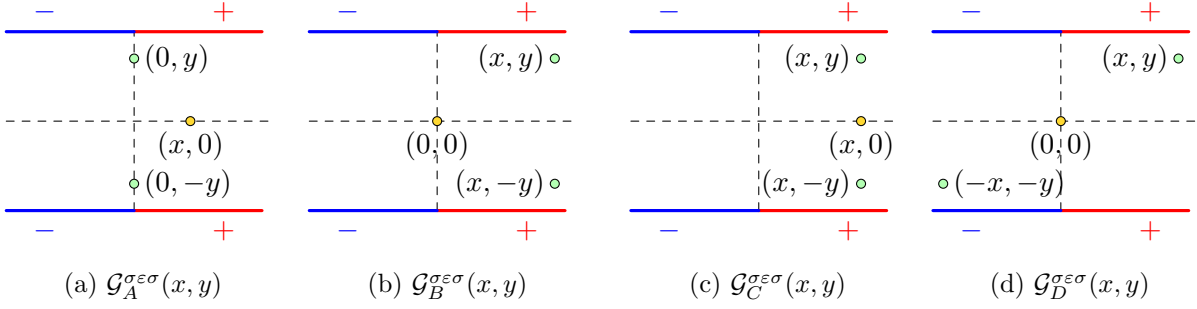

Exact results for mixed correlation function involving two spin fields and one energy density field can be obtained in a straightforward fashion within the probabilistic interpretation \cite{DS_twopoint}. Focusing on the arrangements illustrated in Fig.~ \ref{fig_ssemixed}, the results for the above mentioned mixed correlation functions including corrections at order $O(R^{-1/2})$ are summarized in (\ref{15022021_0717}).

\begin{equation}
\begin{aligned}
\label{15022021_0717}
\mathcal{G}_{A}^{\sigma\varepsilon\sigma}(x,y)/(M^{2} \langle\varepsilon\rangle)& = \frac{2}{\pi} \textrm{sin}^{-1}\left(\rho^{2}\right) + \frac{E}{\sqrt{\pi}\lambda} \textrm{erf}^{2}\left( \frac{\rho\eta}{\sqrt{1-\rho^{2}}} \right) \textrm{e}^{-\eta^{2}} \, , \\
\mathcal{G}_{B}^{\sigma\varepsilon\sigma}(x,y)/(M^{2} \langle\varepsilon\rangle)& = 1 - 8 T\left(\sqrt{2}\chi,\sqrt{\frac{1-\rho^{2}}{1+\rho^{2}}}\right) + \frac{E}{\sqrt{\pi}\lambda} \textrm{erf}^{2}\left( \frac{\chi}{\sqrt{1-\rho^{2}}} \right) \, , \\
\mathcal{G}_{C}^{\sigma\varepsilon\sigma}(x,y)/(M^{2} \langle\varepsilon\rangle)& = 1 - 8 T\left(\sqrt{2}\chi,\sqrt{\frac{1-\rho^{2}}{1+\rho^{2}}}\right) + \frac{E}{\sqrt{\pi}\lambda} \textrm{erf}^{2}\left( \frac{\rho\eta-\chi}{\sqrt{1-\rho^{2}}} \right) \textrm{e}^{-\eta^{2}} \, , \\
\mathcal{G}_{D}^{\sigma\varepsilon\sigma}(x,y)/(M^{2} \langle\varepsilon\rangle)& = -1 + 8 T\left(\sqrt{2}\chi,\sqrt{\frac{1+\rho^{2}}{1-\rho^{2}}}\right) - \frac{E}{\sqrt{\pi}\lambda} \textrm{erf}^{2}\left( \frac{\chi}{\sqrt{1-\rho^{2}}} \right) \, ,
\end{aligned}
\end{equation}
where $E=A_{\varepsilon}^{(0)}/\langle\varepsilon\rangle$. The dependence on the coordinate is encoded in the correlation coefficient $\rho=\sqrt{(1-\tau)/(1+\tau)}$ and the variables $\eta$, $\chi$, which are defined in the main body of the paper. The following clustering relations are easily established:
\begin{equation}
\begin{aligned}
\label{}
\lim_{x \rightarrow + \infty} \mathcal{G}_{A}^{\sigma\varepsilon\sigma}(x,y) & = \mathcal{G}_{\rm i}^{\sigma\sigma}(y,-y) \\
\lim_{x \rightarrow + \infty} \mathcal{G}_{B}^{\sigma\varepsilon\sigma}(x,y) & = \left( 1 + \frac{E}{\sqrt{\pi}\lambda} \right) = - \lim_{x \rightarrow + \infty} \mathcal{G}_{D}^{\sigma\varepsilon\sigma}(x,y) \\
\lim_{x \rightarrow + \infty} \mathcal{G}_{C}^{\sigma\varepsilon\sigma}(x,y) & = M^{2} \langle \varepsilon \rangle \, .
\end{aligned}
\end{equation}

Mixed correlation functions involving two energy density fields and one spin fields at the leading and first subleading order can still be obtained within the probabilistic interpretation.

\bibliographystyle{unsrt}
\bibliography{bibliography}{}

\end{document}